2024/05/09

# A dynamical geography of observed trends in the global ocean

B. Buongiorno Nardelli[1]*, D. Iudicone[2]

[1] Consiglio Nazionale delle Ricerche, Istituto di Scienze Marine, Napoli, Italy
[2] Stazione Zoologica "Anton Dohrn", Napoli, Italy
*Corresponding author. Email: bruno.buongiornonardelli@cnr.it

**Abstract**

Revealing the ongoing changes in ocean dynamics and their impact on marine ecosystems requires the joint analysis of multiple variables. Yet, global observational records only cover a few decades, posing a challenge in the separation of climatic trends from internal dynamical modes. Here, we apply an empirical stochastic model to identify the emergent patterns of trends in six fundamental components of upper ocean physics. We analyze a data-driven reconstruction of the ocean state covering the 1993-2018 period. We found that including temporal derivatives into the state vector enhances the description of the ocean's dynamical system. Once Pacific oscillations are properly accounted for, averaged surface warming appears >60% faster, and a deep reshaping of the seascape is revealed. A clustering of the trend patterns identifies the main factors that drive observed trends in chlorophyll-a concentration. This data-driven approach opens new perspectives in empirical climate modelling.

**Teaser**

A novel analysis of upper ocean physics reveals the factors that shape the seascape evolution in response to global warming.

**MAIN TEXT**

## Introduction

The ocean is a crucial component of Earth's climate system and plays a vital role in regulating the planet's temperature, storing heat and carbon dioxide, and influencing weather patterns and variability. As global warming persists, the oceans are undergoing profound transformations(*1*). To comprehend the implications of these changes and their impact on marine ecosystems, a deeper understanding of their physical drivers is necessary(*2*). This fundamental objective implies a continuous endeavour to improve data analysis techniques(*3–9*) and to exploit innovative data-driven reconstructions(*10–12*) and their use to improve climate models(*13*). Here, we explore the global dynamical patterns that emerge from a multivariate 25-year data-driven ocean state reconstruction. We investigate the upper ocean response to global change, with particular emphasis on the interplay between surface warming and alterations in atmosphere and ocean circulations, and their relation to the hydrological cycle. We assess the extent to which ocean stratification is increasing, and how this modifies the exchanges between the euphotic layer and the deep ocean, looking at how phytoplankton abundance is in turn affected. Our findings highlight that the ocean responses to global warming are more intricate than what can be discerned by examining single variables.

In fact, oceanic trends and variability are generally computed from one single or just a few variables, the most studied being the sea surface temperature (SST)(*1, 14–19*). Resulting temporal and spatial patterns can eventually capture some intrinsic dynamical signal, but do not include sufficient information to describe all the relevant physical components of the system. Moreover, not all techniques are suited to discriminate trends from internal modes of variability, properly accounting for autocorrelation in the time series(*3–9*). Some of the techniques, such as those based on standard Empirical Orthogonal Functions (EOFs), do not necessarily identify physically meaningful patterns, and, by definition, they cannot account for relations that imply wave-like propagating signals or delayed responses. This means also that many of the climatic indexes are not really capturing the full aspects of the underlying dynamical processes(*14, 15*).

A more intriguing method to discover the system dynamics is the empirical fit of a stochastic model to observed time series. A basic assumption of most climatic stochastic models is that the state variables can be separated into low frequency and high frequency components, and that the response to non-linear dynamics decorrelates much faster than the linear dynamics(*4*). In this case, it is possible to describe the system evolution through a linearized dynamical operator plus a white-noise forcing term(*20–23*). Notably, the corresponding system of ordinary differential equations describes an autoregressive (AR) process and its normal modes can be estimated empirically from instantaneous and lagged covariances computed from the observed state vector time series(*20*) (see Methods). This approach is also referred to in the literature as Linear Inverse Modelling (LIM)(*21, 22*). Nevertheless, considering only white-noise forcings means that anthropogenic-induced radiative changes are not explicitly included in the model, and the eigenmodes of the deterministic linear dynamical operator (also known as principal oscillation patterns, POPs[23]) are necessarily affected by neglecting this unknown low-frequency term. The response to the long-term external forcing, however, generally emerges as the least-damped non-oscillating mode(*24*). AR models can be extended to second (or higher) order formulations if the dynamical system displays more complex behaviours. This is equivalent to including both the variables and their time derivatives in the LIM state vector (see Methods).

Here, we used observation-based data(*10–12*) and linear inverse modelling to separate global oceanic trends from interannual-to-decadal oscillations and to provide a dynamical geography of the main drivers of phytoplankton abundance changes over the 1993–2018 period. We estimated POPs to remove oscillating signals from the physical variables, and subsequently identified clusters characterized by consistent trends (see Methods). Finally, we projected chlorophyll-a data (a proxy of phytoplankton abundance) on the trend mode and discussed the physical drivers of observed changes.

## Results

### A novel description of trends from multivariate Linear Inverse Modeling
Our input data include low-pass filtered anomalies (Fig.1) of SST, sea surface salinity (SSS), upper mixed layer depth (MLD), intensity of the horizontal currents (at the surface) and of the vertical exchanges (at 100 m depth), as well as wind stress intensity. These variables provide complementary information: SST integrates the effect of radiative forcing, boundary layer processes and circulation changes; SSS responds to local and non-local hydrological cycle variations (i.e. evaporation (E)/precipitation (P), ice melting, and river discharges), as well as to changes in the ocean circulation and mixing; MLD variations provide an integrated view of the air-sea interaction changes, affecting the vertical distribution and mixing of heat and salt, and the formation and transformation of water masses, a fundamental piece of oceanic 3D circulation(*25*). Surface currents are dominated by the major current systems , while vertical exchanges are mostly driven by the deformation of the mesoscale flow, and to a minor extent by turbulent momentum and buoyancy fluxes(*12*). Finally, the intensity of the downward momentum flux exerted by the atmosphere is included as one of the main elements of the large-scale ocean-atmosphere coupling.

In constructing a LIM, data are usually prefiltered to reduce the impact of noise on the robustness of the computation[19]. Here, we first low-pass filter our multivariate state vector (to meet the scale separation requirement), then we apply EOF decomposition and retain the leading 20 modes as input to our model. These first modes explain ~80% of the original variance and an analysis of the sensitivity to this choice is included in the Methods. We found an autoregressive model of order 2 to best fit our data. Six eigenvalues are real (identifying purely damped modes), and 34 modes occur as complex conjugate pairs, namely describing damped oscillations (see Supplementary text). More than half of the POPs display a quite low percentage of excitation, with values well below 1%. Only one real eigenmode explains a percentage > 1%, displaying the longest damping time among all real modes. Our analysis is restricted to this real eigenmode (hereafter referred to as POP trend mode(*24*)).

**Upper ocean warming patterns and atmospheric circulation changes**
The ocean response to global change reveals an average SST increase of 0.022 ± 0.002 K/yr over the last 25 years. This value agrees with the estimates for recent decades obtained from other datasets(*1*) but is >60% higher than the linear slope estimated from the X-11 trend component over exactly the same period(*26*). The difference is almost completely explained by the negative slope of the linear fit of the Pacific Decadal Variability decay mode amplitude (see Supplementary text). The SST trend pattern, then, slightly differs from what obtained from cyclo-stationary and rotated EOF analyses(*16, 17*), or longer-term linear regressions(*1*). Specifically, while an overall warming is quite evident (Fig.2a), distinct areas of SST decrease are found in both hemispheres. The strong cooling trend observed in the North Atlantic(*27*) is associated with an increasing strength of local westerlies and a significant modification of both the surface salinity field and density gradients, and of the main currents involved in the subpolar gyre and North Atlantic Current system (see Supplementary text).

Noticeably, the two slanted bands of surface cooling found over the North and South Pacific Convergence Zones correspond to the strengthening winds at the edges of the Pacific Hadley cells (Fig.2b). The concomitant weakening of the momentum flux associated with the trade winds in the North Pacific can indeed be explained as a poleward shift of the tropospheric jets(*28*) or, looking at the entire northern hemisphere, as a long-term modification in the patterns of the stationary waves/recurrent ridge systems(*29, 30*). This, in turn, can reflect a long-term expansion of the Hadley(*31*), or a joint modification of the Hadley/Walker circulations(*32*).

Observed patterns show the strongest temperature increase in the northern Pacific. These signals are in fact compatible with the classical wind–evaporation–SST (WES) feedback mechanism, or even with the more recently identified wind-induced turbulence–mixed layer–SST (WIMS) and wind–evaporation–mixed layer–SST (WEMS) feedback mechanisms(*33*). The lower winds along the coasts of the United States lead to a weaker California current (Fig.2f), and to a local slight reduction in the vertical component of the kinetic energy and shallower MLD (Fig.2e). These upwelling-unfavourable conditions contribute to explain the amplified warming offshore California. The simultaneous decrease of the MLD (Fig.2d) is consistent with model projections in presence of decreasing trade winds(*34*).

Concurrent SST decrease and wind stress intensification are observed along the Canary Eastern Boundary Upwelling System (EBUS), the Benguela EBUS, the southern portion of the Humboldt EBUS, and along western Australia. This change originates from an intensification of the trade winds, with the strongest cooling signal observed in the Indian ocean. Consistently, the wind-driven circulation is affected, with a more evident strengthening of the Benguela current and an intensification of the South Pacific subtropical gyre.

**Local and non-local freshwater fluxes and salinity response**
By altering the location and intensity of the low-tropospheric convergence zones, atmospheric circulation changes also impact the E-P balance(*35*, *36*), contributing to shape the SSS trend together with the effect of warming on the hydrological cycle ("dry gets drier, wet gets wetter"(*37*)) and changes in oceanic currents and mixing. Overall, globally averaged salinity change is almost zero (~0.001 g/kg/yr). The strongest signals originate from major rivers' discharge decline recently attributed to anthropogenic factors(*38*). Negative trends appear in the intertropical convergence zone and along most of the Antarctic Circumpolar Current (ACC) (Fig.2c). Salinification appears particularly evident in the tropical southern Atlantic and western Indian oceans, extending further to the South Atlantic Current and to the Agulhas retroflection. Comparing with previous studies(*37*, *39*), the positive SSS trend along the western coasts of the United States, at midlatitudes, as well as the negative trend in the subpolar North Atlantic, reflect a more complex dynamics than a simple response to water cycle amplification(*37*). This implies a direct impact of ocean warming through stratification changes and ice mass loss, as previously speculated from numerical model experiments[39].

**From upper to deeper layers: heat content, barrier layers and intensifying vertical exchanges at the mesoscale**
To investigate how efficiently the surface anomalies are being transported at depth, we estimated the integrated heat and salt content over two layers (0-200 m and 200-1000 m), and the integrated buoyancy anomaly over the first 200 m. Then, we projected each dataset on the POP trend mode (Fig.3-4). The heat content trend in the upper 200 m layer (Fig.3a) is decreasing across a more extensive area in the western Pacific than this identified in the SST trend. This highlights a substantial influence of barrier layers, strengthened by the local E-P fluxes, on regulating surface warming by suppressing deeper mixing(*40*). Moreover, while heat content trend patterns in the two depth ranges look quite similar to each other in the equatorial Pacific Ocean, integrated salinity patterns show a general increase at depth. All these factors lead to a strong increase of the integrated buoyancy anomaly (Fig.4). This stratification change indicates a net reduction of mixing much more clearly than what can be deduced from MLD variations.

A quite different situation is found in the Atlantic Ocean: first of all, all trends appear to be much stronger in the Atlantic with respect to the rest of the global oceans. Then, in contrast with upper layer patterns, both heat and salt content in the 200-1000 m layer display a clear positive trend in the subtropical gyres, with strong maxima along the main western boundary current systems. This is where a marked increase in the vertical exchanges at the mesoscale(*41*) is also observed (Fig.2e), modulating the subtropical mode water formation processes that explain the recent accumulation of excess heat at depth(*42*, *43*).

**MLD changes mostly affect water mass formation areas**
MLD trends provide important information on deep ocean ventilation processes. In the North Atlantic (see also Supplementary text), the increase in wind stress, together with the modification of the complex sub-polar gyre circulation patterns, drive a general increase in MLD, except for the areas of dense water formation in the Labrador Sea. Conversely, even if an intensifying wind stress is found along most of the ACC, the general MLD tendency there is to get shallower south of the mean position of the Sub-Antarctic Front. More precisely, the MLD trend along the ACC derives from a concomitant surface salinity reduction and temperature increase (impacting the surface density), but also reflects a latitudinal translation of the Sub-Antarctic Front. This shift, identified as an alternated positive/negative thin band in the horizontal current intensity trend (Fig.2f), is only noticeable in specific areas: a southward displacement at the Southeast Indian Ridge, and a northward displacement between around the international date line and 110°W. Meanwhile, the vertical component of the kinetic energy at 100 m depth is also slightly increasing. Long-term MLD and mesoscale activity trends are thus likely driving modifications in the water mass subduction in the areas where most of the Antarctic Intermediate Water and Sub-Antarctic Mode Water are formed(*42*, *44*).

**Trend clustering reveals emergent dynamical patterns**
To discuss the physical mechanisms driving marine phytoplankton changes, we first clustered the trend patterns observed in our six physical variables (Fig.5) and then focused on chlorophyll-a response seen within each cluster. Chlorophyll-a trend was obtained by projecting satellite estimates on the POP mode (see Methods). Corresponding pattern (Fig.6) differs from previous estimates obtained through simpler linear regression techniques(*45*), which, by construction, cannot handle autocorrelation in the time series properly(*46*).

We applied two clustering algorithms (based on *K-means* and *K-medoids*, see Methods) and tested different choices of the target number of clusters. In the following, we focus on the 9 clusters obtained with *K-means*, as they provide a reasonable level of granularity and are characterized by distinct trends in few physical variables. Different choices lead to relatively minor differences and have no major impact on the discussion (see Methods). Only in some cases the dynamical regimes show a clear/unique sign response in chlorophyll-a.

The strongest surface warming is associated with a marked freshening and decrease in the wind stress intensity. The corresponding cluster (blue areas in Fig.5a) covers approximately one fifth of the ocean surface and is predominantly located around the Equator, particularly in the Pacific. This cluster almost perfectly identifies the area where the ocean stratification (as measured by the integrated buoyancy anomaly shown in Fig.4) is experiencing a robust increase. This pattern cannot be identified by looking separately at SST or MLD trends. Here, a significant decline in chlorophyll-a is observed, (Fig.5b and Fig.6), likely due to the inhibited resupply of nutrients caused by the enhanced stratification. A moderate temperature rise accompanied by a clear salinity increase is associated with the second largest cluster (light olive color in Fig.5), accounting for 27.9% of the area. These conditions dominate the western Indian Ocean, the Mediterranean Sea and the equatorial Atlantic. In the Pacific Ocean, this cluster identifies very clearly the California EBUS and the northern portion of the Humboldt EBUS (along the coasts of Peru), where temperature rise is related to a decline in the upwelling. Weaker upwelling is driving, in turn, a tangible chlorophyll-a decrease (Fig.5b and Fig.6). Indeed, its projected trend shows a seamless negative patch in the entire north-eastern Pacific, but the mechanisms driving this decline are thus quite different depending on the area considered. The highest percentage of the ocean surface (36.2%, violet cluster) is associated with a quite mild surface warming and slightly intensifying wind stress, the latter supporting the increase in chlorophyll-a due to the intensifying EBUS in the Atlantic and along the Southern coasts of Chile. The fourth and fifth clusters (cyan and red areas in Fig.5, occupying 7.7% and 5.7% of the domain, respectively) show opposite behaviours: the former is characterized by a significant shoaling of the upper mixed layer associated with surface warming. It basically identifies changes occurring at the northern boundary of the North-Atlantic subtropical gyre, and at the southern edges of the Antarctic Circumpolar Current (ACC) in the central sectors of the Pacific and Indian Oceans. Conversely, the latter identifies areas subject to enhanced mixing, generally coinciding with cooling and freshening of the surface waters and increasing wind stress. This cluster covers most of the mode water formation areas in the subpolar North Atlantic, the Agulhas current system and a wide sector of the south-eastern Indian Ocean. The chlorophyll-a signal inside these clusters is very weak, but consistently displays an opposite sign with respect to the MLD trend. This may be attributed to the reduced light availability as mixing extends to greater depths. The sixth grouping (orange cluster in Fig.5, covering 0.9% of the area) highlights a marked intensification of the vertical exchanges, a moderate increase in the intensity of the horizontal surface currents and a slight increase in MLD, all elements pointing to a strengthening of the mesoscale activity. The seventh cluster presents an almost specular behaviour, describing the reduction of the Kuroshio current width and its downstream intensification(*47*). Phytoplankton response is consistent: weaker vertical exchanges lead to lower chlorophyll-a and vice versa(*48*). The eighth and nine clusters instead reflect the decrease in the freshwater flux from the Mississippi, Amazon and Rio de la Plata rivers, distinguishing between main river outflows (green cluster) and transitional waters. Chlorophyll-a here evidences a negative impact of reduced runoff, though potentially affected by the lower accuracy of chlorophyll retrieval algorithms in coastal waters.

**Discussion**

Our findings are undeniably affected by several limitations, one of the most relevant being the reduced length of observational time series. Still, the fact that the ocean dynamical system is better described by including multiple variables and their time derivatives in the state vector represents by itself a relevant conceptual advance for empirical climate modelling. By introducing a novel dynamical geography of seascape change, we discern the local physical drivers of the observed trends in chlorophyll concentration beyond simple temperature-driven stratification changes(*49*). Our approach could represent a "higher order" reference for the validation of numerical climate models. This information may foster new investigations of the impact of climate change on marine ecosystems, e.g. by comparing our dynamical regimes with recently identified plankton genomic

biogeographies(*50*), and can be used to define more effective sampling strategies for ocean monitoring and ecosystem preservation(*51*).

**Materials and Methods**

### Data collection and pre-processing

Sea surface temperature, salinity, and mixed layer depth data have been extracted from the European Copernicus Marine Service ARMOR3D product (https://doi.org/10.48670/moi-00052). ARMOR3D is based on a combination of state-of-the-art satellite-derived surface temperature, surface salinity and altimetric data and in situ observations(*10*). This dataset is provided weekly on a ¼° resolution grid and covers a depth range of up to 5500 m. It contains 50 vertical levels, with higher resolution in the upper 1500 m of the water column (between 5 and 100 m) and lower resolution below (between 250 and 500 m).

2D and 3D ocean currents have been extracted from European Copernicus Marine Service Globcurrent(*11*) and OMEGA3D(*12*) products. Copernicus-Globcurrent provides global total velocity fields at 0 m and 15 m at ¼° spatial resolution. It is obtained by combining satellite altimeter-derived geostrophic surface currents and modelled Ekman currents estimated from ECMWF ERA5 wind stress (https://doi.org/10.48670/moi-00050).

OMEGA3D is based on ARMOR3D data and covers the period from 1993 to 2018 (https://doi.org/10.25423/cmcc/multiobs_glo_phy_w_rep_015_007). This dataset provides 3D fields of both vertical and horizontal quasi-geostrophic currents, which are estimated through a diagnostic model based on a diabatic version of the Omega equation. The model incorporates different processes such as kinematic deformation, turbulent buoyancy and momentum fluxes, and takes in input the surface forcing estimated from ERA-Interim atmospheric re-analysis (https://doi.org/10.21957/pocnex23c6). The data are provided weekly on the same horizontal grid as ARMOR3D, except for the equatorial band, where quasi-geostrophic diagnostics cannot be applied. Full details on the methodology and data set validation can be found in the literature[12]. The intensity of the horizontal and vertical currents used here are both defined as the square root of corresponding kinetic energy components.

Wind stress intensity is taken from the European Centre for Medium-Range Weather Forecasts (ECMWF) ERA5 global atmospheric reanalysis (https://doi.org/10.24381/cds.adbb2d47). The ERA5 system utilizes the ECMWF Earth System model IFS, cycle 41r2, which incorporates multiple observations of upper-air atmospheric variables, including satellite radiances, temperature, wind vectors, specific humidity, and ozone. We sampled the ERA5 mean daily wind stress data taking the same weekly dates.

Ocean Chlorophyll-a concentration data are taken from the European Copernicus Marine Service GlobColour daily, 4 km resolution interpolated data (https://doi.org/10.48670/moi-00281). The data sources for this product include SeaWiFS, MODIS, MERIS, VIIRS-SNPP, JPSS1, OLCI-S3A, and OLCI-S3B. We sampled the Copernicus-GlobColour data taking the same weekly dates as for ARMOR3D and OMEGA3D products and mapped the data on the same grid.

# Linear inverse modelling, Principal Oscillation Patterns and Autoregressive processes

A Linear Inverse Model (LIM) assumes that the system's evolution can be explained as a linear process forced by stochastic (white) noise. The evolution of the (multivariate) system state variable $x$ is thus driven by a deterministic time-independent linear dynamical operator $L$ and by a component $\xi$ which represents some random fluctuations that may eventually be spatially correlated, but not correlated over time. The corresponding stochastic differential equation (SDE) is:

$$\frac{d\boldsymbol{x}}{dt} = \boldsymbol{L}\boldsymbol{x} + \boldsymbol{\xi}. \qquad (1)$$

The most likely evolution of $x$ after a temporal lag $\tau$ is described by an autoregressive process:

$$\widehat{\boldsymbol{x}}(t+\tau) = exp(\boldsymbol{L}\tau)\,\boldsymbol{x}(t) = \boldsymbol{G}(\tau)\boldsymbol{x}(t), \qquad (2)$$

where the propagator $G$, also called Green's function, can be estimated from observed instantaneous and lagged covariances:

$$\boldsymbol{G}(\tau) = \boldsymbol{C}_\tau \boldsymbol{C}_0 \qquad (3)$$

The eigenmodes of $L$ are found by solving:

$$\boldsymbol{L}\boldsymbol{U} = \boldsymbol{P}\boldsymbol{U} \qquad (4)$$

and can be either real or complex (in which case they appear as conjugate pairs), where $\Lambda$ identifies the eigenvalues ($\lambda_k$). These eigenmodes provide the empirical dynamical modes/Principal Oscillation Patterns of the system ($\boldsymbol{u}_k$):

$$\boldsymbol{x}(t) = \sum_{k=1}^{m} Z_k(t)\boldsymbol{u}_k \qquad (5)$$

each characterized by a specific period:

$$T_k = \frac{2\pi}{|arg(\lambda_k)|} \qquad (6)$$

and damping time:

$$\tau_k = \frac{-1}{log(|\lambda_k|)} \qquad (7)$$

The temporal amplitudes of the POP modes, the coefficients $Z_k(t)$, are computed by projecting the original data on the eigenvectors of the linear operator adjoint, defined by:

$$\boldsymbol{L^H V = V \Lambda^*} \qquad (8)$$

$$Z_k(t) = \boldsymbol{x}(t)^T \boldsymbol{v}_k \qquad (9)$$

where the superscripts H, * and T indicate the transpose conjugate, the conjugate and the transpose, respectively.

Complex POPs are able to reveal lagged feedback (both local and remote) between state vector elements, and provide damped/oscillating solutions, thus allowing to efficiently describe phenomena characterized by evolving spatial patterns and non-trivial dynamical relations. The variance of the k-th POP amplitude, also called "excitation", measures the dynamical relevance of corresponding eigenmode (provided a proper normalization is preliminary carried out(*21*)). We express it here as a percentage by normalizing the variance of each mode amplitude by the total variance.

The discretized version of equation (1) actually describes a discrete multivariate autoregressive process of order 1:

$$\boldsymbol{x}_t = \mathbf{A} \boldsymbol{x}_{t-1} + \boldsymbol{\varepsilon}_t \qquad (10)$$

Models of higher order, AR(p), include state vectors values up to *p* previous time steps and they can always be transformed in AR(1) models by augmenting the state vector dimensions as described in(*21*). For an AR(2) model, this is in fact equivalent, in the continuous formulation, to including both the original state variable and its first order derivative in the state vector(*52*), basically in the same way as a second order SDE can be reduced to a system of first order SDEs(*53*). For further details on the relation between Linear Inverse Modelling, Principal Oscillation Patterns and Autoregressive modelling, the readers can refer to several papers and books listed in our reference list(*20, 21, 54, 55*). POP analysis has thus been carried out here by testing different Autoregressive (AR) model configurations. AR modelling has been carried out using the *ARfit* package(*56*).

**AR input data filtering and dimensionality reduction**

POP modelling could in principle be applied directly to the original data, the only methodological choice being then related to the order of the AR model. However, AR decomposition can be problematic in presence of excessively noisy data. It can also be quite costly from the computational point of view. As such, it is common practice to reduce the dimensionality of the problem and to filter noise by preliminary estimating the Empirical Orthogonal Function decomposition (EOF, also known as Principal Component

Analysis, PCA) of the state vector time series and keeping only a subset of leading Principal Components (PCs) for subsequent analyses.

EOF analysis is likely the most widely used technique to identify variability modes. EOF/PCA decomposes a given time series of spatial data into a linear combination of modes, obtained as the eigenvectors of the covariance matrix of the data set. These basis functions consist of fixed spatial patterns and the projection of the original dataset on these vectors provides temporal amplitudes that are also known as principal components. A few modes are generally sufficient to describe a high percentage of the global variance, which makes EOF quite useful to simplify the analysis of large datasets. To our knowledge, all applications of AR models in the climate community adopt this EOF truncation strategy(*14*, *19*, *55*, *57*).

Here, we have computed the multivariate EOFs including in the input vector the six dynamical variables identified above (SST, SSS, MLD, surface horizontal current intensity, root square of vertical component of the kinetic energy at 100 m depth, wind stress intensity). As a preliminary step, each variable has been normalized before entering the analysis. We have then low-pass filtered all timeseries to keep only slowly-varying signals (this is a prerequisite to build a LIM of the climate system(*4*)). This filtering has been carried out on each time series (at individual grid-points) through a double moving-average smoothing, in order to reduce the spectral leakage while keeping the computation efficient (first taking a 53 weeks window and then a 39 weeks window). The weekly time series has been finally subsampled to monthly resolution.

Ideally, if data were not corrupted by noise, we could have used all PCs as input to the AR modelling, as EOFs represent a complete basis set for the original data, already achieving a significant reduction of the dimensionality of the problem. However, observation-based data do include noisy components that might preclude a robust estimation of the AR parameters. The number of PCA/EOF modes to retain should thus preserve as much as possible the signal variance and remove only PCs that can be clearly associated with noise. This step is further complicated by the sampling errors incurred in estimating the EOF shapes, though.

A common method to decide whether an EOF mode can be correctly interpreted is to compare the difference between neighbouring eigenvalues with corresponding sampling error. When a group of eigenvalues lie within the confidence interval, they form an "effectively degenerate multiplet" and interpretation becomes difficult because the signal associated with a specific process may be mixed/splitted among different modes. This is commonly known as the North's rule of thumb(*58*). However, if the objective is not to analyse the shape of the EOFs per se, mixed multiplet members can still add relevant information to a truncated EOF reconstruction, provided one takes care that the truncation point does not fall in the middle of an "effective multiplet". Thus, additional assumptions about the noise must be made. A simple approach, sometimes called Kaiser's rule, is to retain only those PCs whose eigenvalues lie above a threshold value (a conservative choice being the average of all eigenvalues)(*59*). Given the values reported in Fig.S1, in our case we should thus not extend the EOF decomposition beyond mode 34.

The number of PCs considered, however, may influence the estimated optimal AR model order and its stability, as residual noise and/or improperly sampled physical processes can still contaminate the calculations if too many PCs are included. Therefore, we had to test and assess the impact of EOF truncation on the stability of our autoregressive models. We started from 34 PCs and found that AR models (of up to order 5) are not dynamically stable, namely some roots of the characteristic equation are not outside of the unit circle.

In such cases, the estimates obtained using least squares may not converge to the true parameters of the model, and the model parameters are not reliable.

We thus gradually reduced the number of PCs in input until we could obtain a stable AR model. This occurs when less than 23 PCs are considered, and only if an AR model of order >=2 is adopted. To be sure that our choice of the number of PCs to retain is not introducing artefacts by truncating an effective multiplet, we thus focused on 20 PCs and compared resulting AR eigenmodes with those obtained considering 18 to 22 PCs. Using 20 PCs guarantees that more than 80% of the variance is kept in the subsequent analysis. Both trend mode spatial patterns and temporal coefficients are almost identical whatever the configuration considered (not shown). The slight differences found were then used to associate an uncertainty to the fitted linear trend (given as the standard deviation of the 5 values).

**AR model robustness and model order selection**

AR models intrinsically assume the noise vectors (which correspond to the stochastic forcing of the system) to be uncorrelated, and the uncorrelatedness of the noise vectors is invoked in the derivation of the least squares estimator.

After fitting an AR model to empirical data, this assumption can be easily verified by looking at the autocorrelation function (ACF) of the model residuals (i.e. the lagged autocorrelation of the difference between input data and modelled time series). If we fit our data to an AR(1) model and look at corresponding ACF (shown in Fig.S2), we clearly see that residuals display high levels of autocorrelation at several lags. This demonstrates that an AR(1) model is not sufficient to describe the processes we are trying to model. Conversely, residual ACFs immediately drop and stay below (or very close to) the approximate 95% confidence limits when an AR(2) model is used (see Fig.S3). The robustness and parsimony of the choice of model order is then confirmed by Bayesian Information Criterion, as using higher order AR models would imply adding more parameters without adding significant improvements. These results are robust in the range of PCs considered (not shown).

**POP analysis robustness**

As fully discussed in(*21*), only approximate confidence intervals can be empirically estimated for the coefficients of an AR model, based on the asymptotic distribution of the least squares estimator.

Establishing confidence intervals for the eigenmodes and their periods and damping times is further complicated by the fact that these quantities are nonlinear functions of the AR coefficients, so that additional approximations are needed. In fact, even for large samples, the approximate confidence intervals for the eigenmodes, periods, and damping times are less accurate than the confidence intervals for the AR parameters themselves. In practice, these approximate 95% margins of error are very rough indicators of the estimation errors, especially in presence of relatively small sample size, so that they are almost never discussed in practical applications to climate data(*55, 57*). Both periods and damping times and related approximate confidence intervals are presented in the Supplementary (see also Table S1).

**Clustering algorithms and number of clusters selection**

We tested two different clustering algorithms: *K-means* and *CLARA* (Clustering Large Applications(*60*)), which is a numerically efficient implementation of *K-medoids*(*61*), both based on *Scikit-learn*(*62*).

*K-means* works by calculating the mean of data points within each cluster and moving the centroid to that mean. It minimises the sum of squared distances between data points and their respective cluster centroids. Here it is initialized with the *greedy k-means++* strategy(*63*). Instead of using the mean, *K-medoids* uses the medoid, which is the most centrally located point in a cluster. The medoid is chosen as the data point that has the smallest average dissimilarity to all the other points in the cluster. *CLARA* is designed to work well with large datasets by sampling subsets of the data to find representative medoids. The metric used is Euclidean distance in both *K-means* and *CLARA* algorithms. We have assessed the impact of the choice of the clustering algorithm on our results by comparing the results obtained with *K-means* (Fig. 3) with those obtained with *CLARA*. The classifications present minor differences that are consistent with the fact that *CLARA* is by construction less affected by extreme values and the details have been included in a specific section in the Supplementary.

Both algorithms require as input parameter the target number of clusters. We used four different approaches to evaluate which number of clusters would better serve our objectives. They give contrasting results (see Fig.S4-S5). The approaches are: the "elbow" criterion, the Davies-Bouldin score, the "silhouette" score and Bayesian Information Criterion (BIC).

In our case, BIC would always suggest a very high number of clusters, which would not be useful for our objective, that is to provide a synthetic view of the main multivariate trend patterns identified by our autoregressive model. It would not be useful to this aim having tens of clusters, especially if these identify extremely small areas and/or only slightly differing dynamical regimes. However, BIC is based on standard likelihood theory, which does not apply to the partitioning clustering algorithms we have considered, so it is likely not the best choice to obtain meaningful and/or easily interpretable results(*64*).

The Davis-Bouldin Index (DBI) is a measure of cluster separation and compactness(*65*). It calculates the ratio of the average intra-cluster distance to the distance between cluster centroids. Lower DBI values indicate better clustering, where clusters are more separated and compact. In our calculations, DBI shows a sharp decline anticipating a sort of minimum plateau for both *K-means* and *CLARA*. The plateau is reached between 8 and 11 clusters, continuing to decrease up to 25 in *CLARA*, but starting to increase after 14 in *K-means*.

The silhouette score is another way to measure how similar an object is to its own cluster compared to other clusters(*66*). In both *K-means* and *K-medoids*, the silhouette score provides a quantitative measure of how well the data is clustered. A score close to +1 indicates that the data point is well-clustered, a score around 0 indicates overlapping clusters, and a score close to -1 indicates that the data point may have been assigned to the wrong cluster. In our tests, the silhouette score oscillates around relatively low values, never exceeding 0.15, and dropping markedly over 15 clusters.

The elbow criterion is based on the principle that as the number of clusters increases, the within-cluster sum of squares (WCSS) decreases. It looks for the point where the rate of decrease in WCSS slows down abruptly, forming an "elbow" shape in the plot. The plots of WCSS would suggest an optimal number of clusters between 8 and 13.

Finally, we have thus selected to analyse 9 clusters, as this is the number of clusters that is better informing on general patterns without introducing excessive granularity (basically following the elbow criterion). Similar "kind-of-subjective" choices are common in the literature dedicated to global change(*50*).

**References**
1. C. Garcia-Soto *et al.*, An Overview of Ocean Climate Change Indicators: Sea Surface Temperature, Ocean Heat Content, Ocean pH, Dissolved Oxygen Concentration, Arctic Sea Ice Extent, Thickness and Volume, Sea Level and Strength of the AMOC (Atlantic Meridional Overturning Circulation). *Front. Mar. Sci.* **8** (2021), doi:10.3389/fmars.2021.642372.
2. Bindoff et al., Chapter 5 - Changing Ocean, Marine Ecosystems, and Dependent Communities In *The Ocean and Cryosphere in a Changing Climate* (Cambridge University Press, 2022; https://www.cambridge.org/core/product/identifier/9781009157964%23c5/type/book_part), pp. 447–588.
3. S. Sippel *et al.*, Robust detection of forced warming in the presence of potentially large climate variability. *Sci. Adv.* **7**, 4429–4451 (2021).
4. K. Hasselmann, Stochastic climate models: Part I. Theory. *Tellus A Dyn. Meteorol. Oceanogr.* **28**, 473 (1976).
5. C. Penland, Random forcing and forecasting using principal oscillation pattern analysis. *Mon. Weather Rev.* **117**, 2165–2185 (1989).
6. A. Hannachi, I. T. Jolliffe, D. B. Stephenson, Empirical orthogonal functions and related techniques in atmospheric science: A review. *Int. J. Climatol.* **27**, 1119–1152 (2007).
7. C. L. E. Franzke, R. Blender, T. J. O'Kane, V. Lembo, Stochastic Methods and Complexity Science in Climate Research and Modeling. *Front. Phys.* **10**, 1–12 (2022).
8. M. Ghil, V. Lucarini, The physics of climate variability and climate change. *Rev. Mod. Phys.* **92**, 35002 (2020).
9. H. Von Storch, F. Zwiers, *Statistical Analysis in Climate Research* (2001; http://books.google.com/books?hl=en&lr=&id=_VHxE26QvXgC&oi=fnd&pg=PP1&dq=Statistical+Analysis+in+Climate+Research&ots=Ejo_v5zvbu&sig=DtZUVbB_DxRQoxQa9uCwN8AsL4A).
10. S. Guinehut, a.-L. Dhomps, G. Larnicol, P.-Y. Le Traon, High resolution 3-D temperature and salinity fields derived from in situ and satellite observations. *Ocean Sci.* **8**, 845–857 (2012).
11. M. Rio, S. Mulet, N. Picot, Beyond GOCE for the ocean circulation estimate: Synergetic use of altimetry, gravimetry, and in situ data provides new insight into geostrophic and Ekman currents. *Geophys. Res. Lett.* **41**, 8918–8925 (2014).
12. B. Buongiorno Nardelli, A multi-year time series of observation-based 3D horizontal and vertical quasi-geostrophic global ocean currents. *Earth Syst. Sci. Data*. **12**, 1711–1723 (2020).
13. V. Eyring *et al.*, Taking climate model evaluation to the next level. *Nat. Clim. Chang.* **9**, 102–110 (2019).
14. C. Penland, L. Matrosova, Studies of El Niño and Interdecadal Variability in Tropical Sea Surface Temperatures Using a Nonnormal Filter. *Source J. Clim.* **19**, 5796–5815 (2006).
15. G. P. Compo, P. D. Sardeshmukh, Removing ENSO-related variations from the climate record. *J. Clim.* **23**, 1957–1978 (2010).
16. S. R. Yeo, S. W. Yeh, K. Y. Kim, W. M. Kim, The role of low-frequency variation in the manifestation of warming trend and ENSO amplitude. *Clim. Dyn.* **49**, 1197–1213 (2017).

17. X. Chen, K. K. Tung, Global-mean surface temperature variability: space–time perspective from rotated EOFs. *Clim. Dyn.* **51**, 1719–1732 (2018).
18. M. A. Alexander, S. Shin, D. S. Battisti, The Influence of the Trend, Basin Interactions, and Ocean Dynamics on Tropical Ocean Prediction. *Geophys. Res. Lett.* **49** (2022), doi:10.1029/2021GL096120.
19. E. Di Lorenzo et al., Modes and Mechanisms of Pacific Decadal-Scale Variability. *Ann. Rev. Mar. Sci.* **15**, 249–275 (2023).
20. A. Hannachi, *Patterns Identification and Data Mining in Weather and Climate* (Springer International Publishing, Cham, 2021; https://link.springer.com/10.1007/978-3-030-67073-3), *Springer Atmospheric Sciences*.
21. A. Neumaier, T. Schneider, Estimation of parameters and eigenmodes of multivariate autoregressive models. *ACM Trans. Math. Softw.* **27**, 27–57 (2001).
22. C. Penland, P. D. Sardeshmukh, The Optimal Growth of Tropical Sea Surface Temperature Anomalies. *J. Clim.* **8**, 1999–2024 (1995).
23. K. Hasselmann, PIPs and POPs: the reduction of complex dynamical systems using principal interaction and oscillation patterns. *J. Geophys. Res.* **93** (1988), doi:10.1029/jd093id09p11015.
24. C. Frankignoul, G. Gastineau, Y.-O. Kwon, Estimation of the SST Response to Anthropogenic and External Forcing and Its Impact on the Atlantic Multidecadal Oscillation and the Pacific Decadal Oscillation. *J. Clim.* **30**, 9871–9895 (2017).
25. S. Groeskamp et al., The Water Mass Transformation Framework for Ocean Physics and Biogeochemistry. *Ann. Rev. Mar. Sci.*, 1–35 (2018).
26. K. von Schuckmann et al., Copernicus Marine Service Ocean State Report, Issue 4. *J. Oper. Oceanogr.* **13**, S1–S172 (2020).
27. S. A. Josey et al., The recent atlantic cold anomaly: Causes, consequences, and related phenomena. *Ann. Rev. Mar. Sci.* **10**, 475–501 (2018).
28. A. J. Simmons, Trends in the tropospheric general circulation from 1979 to 2022. *Weather Clim. Dyn.* **3**, 777–809 (2022).
29. Y. C. Liang, J. Y. Yu, E. S. Saltzman, F. Wang, Linking the tropical Northern Hemisphere pattern to the pacific warm blob and Atlantic cold blob. *J. Clim.* **30**, 9041–9057 (2017).
30. R. C. J. Wills, R. H. White, X. J. Levine, Northern Hemisphere Stationary Waves in a Changing Climate. *Curr. Clim. Chang. Reports*. **5** (2019), pp. 372–389.
31. T. Xian et al., Is Hadley Cell Expanding? *Atmosphere (Basel)*. **12**, 1699 (2021).
32. K. S. Yun, A. Timmermann, M. Stuecker, Synchronized spatial shifts of Hadley and Walker circulations. *Earth Syst. Dyn.* **12**, 121–132 (2021).
33. T. Kataoka, M. Kimoto, M. Watanabe, H. Tatebe, Wind-mixed layer-SST feedbacks in a tropical air-sea coupled system: Application to the Atlantic. *J. Clim.* **32**, 3865–3881 (2019).
34. N. Maher, M. H. England, A. Sen Gupta, P. Spence, Role of Pacific trade winds in driving ocean temperatures during the recent slowdown and projections under a wind trend reversal. *Clim. Dyn.* **51**, 321–336 (2018).
35. M. P. Byrne, A. G. Pendergrass, A. D. Rapp, K. R. Wodzicki, Response of the Intertropical Convergence Zone to Climate Change: Location, Width, and Strength. *Curr. Clim. Chang. Reports*. **4**, 355–370 (2018).
36. J. R. Brown et al., South Pacific Convergence Zone dynamics, variability and impacts in a changing climate. *Nat. Rev. Earth Environ.* **1**, 530–543 (2020).
37. P. J. Durack, S. E. Wijffels, R. J. Matear, Ocean Salinities Reveal Strong Global Water Cycle Intensification During 1950 to 2000. *Science (80-. ).* **336**, 455–458 (2012).
38. L. Gudmundsson et al., Globally observed trends in mean and extreme river flow attributed to climate change. *Science (80-. ).* **371**, 1159–1162 (2021).
39. J. D. Zika et al., Improved estimates of water cycle change from ocean salinity: the key

role of ocean warming. *Environ. Res. Lett.* **13**, 074036 (2018).
40. L. Wang, F. Xu, Decadal variability and trends of oceanic barrier layers in tropical Pacific. *Ocean Dyn.* **68**, 1155–1168 (2018).
41. J. Martínez-Moreno et al., Global changes in oceanic mesoscale currents over the satellite altimetry record. *Nat. Clim. Chang.* **11**, 397–403 (2021).
42. Z. Li, M. H. England, S. Groeskamp, Recent acceleration in global ocean heat accumulation by mode and intermediate waters. *Nat. Commun.* **14**, 6888 (2023).
43. B. Gan et al., North Atlantic subtropical mode water formation controlled by Gulf Stream fronts. *Natl. Sci. Rev.* **10** (2023), doi:10.1093/nsr/nwad133.
44. T. Qu, S. Gao, R. A. Fine, Variability of the Sub-Antarctic Mode Water Subduction Rate During the Argo Period. *Geophys. Res. Lett.* **47** (2020), doi:10.1029/2020GL088248.
45. K. von Schuckmann et al., Copernicus Marine Service Ocean State Report, Issue 3. *J. Oper. Oceanogr.* **12**, S1–S123 (2019).
46. B. B. Cael, K. Bisson, E. Boss, S. Dutkiewicz, S. Henson, Global climate-change trends detected in indicators of ocean ecology. *Nature.* **619**, 551–554 (2023).
47. Y. Zhang, Z. Zhang, D. Chen, B. Qiu, W. Wang, Strengthening of the Kuroshio current by intensifying tropical cyclones. *Science (80-. ).* **368**, 988–993 (2020).
48. M. Lévy et al., The Impact of Fine-Scale Currents on Biogeochemical Cycles in a Changing Ocean. *Ann. Rev. Mar. Sci.* **16**, 1–25 (2024).
49. M. J. Behrenfeld et al., Climate-driven trends in contemporary ocean productivity. **444**, 752–755 (2006).
50. P. Frémont et al., Restructuring of plankton genomic biogeography in the surface ocean under climate change. *Nat. Clim. Chang.* **12**, 393–401 (2022).
51. E. Boss et al., Recommendations for Plankton Measurements on OceanSITES Moorings With Relevance to Other Observing Sites. *Front. Mar. Sci.* **9**, 1–16 (2022).
52. S. M. Pandit, S. M. Wu, Unique Estimates of the Parameters of a Continuous Stationary Stochastic Process. *Biometrika.* **62**, 497 (1975).
53. K. Burrage, I. Lenane, G. Lythe, Numerical Methods for Second-Order Stochastic Differential Equations. *SIAM J. Sci. Comput.* **29**, 245–264 (2007).
54. C. Penland, L. Matrosova, Prediction of Tropical Atlantic Sea Surface Temperatures Using Linear Inverse Modeling. *J. Clim.* **11**, 483–496 (1998).
55. H. Von Storch, G. Burger, R. Schnur, J. S. Von Storch, Principal oscillation patterns: a review. *J. Clim.* **8**, 377–400 (1995).
56. T. Schneider, A. Neumaier, Algorithm 808: ARfit---a matlab package for the estimation of parameters and eigenmodes of multivariate autoregressive models. *ACM Trans. Math. Softw.* **27**, 58–65 (2001).
57. W. J. Crawford, A. M. Moore, J. Fiechter, C. A. Edwards, A Principal Oscillation Pattern Analysis of the Circulation Variability in the California Current System associated With the El Niño Southern Oscillation. *J. Geophys. Res. Ocean.* **124**, 8241–8256 (2019).
58. G. North, T. Bell, Sampling Errors in the Estimation of Empirical Orthogonal Funtions. *Mon. Weather ….* **110** (1982), pp. 699–706.
59. I. T. Jolliffe, *Principal Component Analysis* (Springer-Verlag, New York, Second., 2002; https://www.taylorfrancis.com/books/9781351206341/chapters/10.4324/9781351206358-1), *Springer Series in Statistics*.
60. L. Kaufman, P. J. Rousseeuw, (1990; https://onlinelibrary.wiley.com/doi/10.1002/9780470316801.ch3), pp. 126–163.
61. L. Kaufman, P. J. Rousseeuw, in *Finding Groups in Data: An Introduction to Cluster Analysis* (John Wiley & Sons, Inc, 1990; https://onlinelibrary.wiley.com/doi/10.1002/9780470316801.ch2), pp. 68–125.
62. F. Pedregosa et al., Scikit-learn: Machine Learning in Python. *J. Mach. Learn. Res.* **12**,

2825–2830 (2011).
63. D. Arthur, S. Vassilvitskii, K-means++: The advantages of careful seeding. *Proc. Annu. ACM-SIAM Symp. Discret. Algorithms*. **07-09-Janu**, 1027–1035 (2007).
64. C. Hennig, Clustering strategy and method selection. *Handb. Clust. Anal.*, 703–730 (2015).
65. D. L. Davies, D. W. Bouldin, A Cluster Separation Measure. *IEEE Trans. Pattern Anal. Mach. Intell.* **PAMI-1**, 224–227 (1979).
66. P. J. Rousseeuw, Silhouettes: A graphical aid to the interpretation and validation of cluster analysis. *J. Comput. Appl. Math.* **20**, 53–65 (1987).
67. L. Caesar, G. D. McCarthy, D. J. R. Thornalley, N. Cahill, S. Rahmstorf, Current Atlantic Meridional Overturning Circulation weakest in last millennium. *Nat. Geosci.* **14**, 118–120 (2021).
68. C. He *et al.*, A North Atlantic Warming Hole Without Ocean Circulation. *Geophys. Res. Lett.* **49** (2022), doi:10.1029/2022GL100420.
69. L. Li, M. S. Lozier, F. Li, Century-long cooling trend in subpolar North Atlantic forced by atmosphere: an alternative explanation. *Clim. Dyn.* **58**, 2249–2267 (2022).
70. N. P. Holliday *et al.*, Ocean circulation causes the largest freshening event for 120 years in eastern subpolar North Atlantic (2020), doi:10.1038/s41467-020-14474-y.
71. T. Petit, M. S. Lozier, S. A. Josey, S. A. Cunningham, Role of air-sea fluxes and ocean surface density in the production of deep waters in the eastern subpolar gyre of the North Atlantic. *Ocean Sci.* **17**, 1353–1365 (2021).
72. J. Sun, M. Latif, W. Park, Subpolar gyre–AMOC–atmosphere interactions on multidecadal timescales in a version of the Kiel climate model. *J. Clim.* **34**, 6583–6602 (2021).

**Acknowledgments**

The data-driven reconstructions used here were produced and made freely available by the European Union Copernicus Marine Service (https://marine.copernicus.eu) and Copernicus Climate Change Service (https://climate.copernicus.eu). We thank prof. Tapio Schneider for sharing his tool for autoregressive model fitting.

**Funding:** Atlantic ECOsystems assessment, forecasting & sustainability (AtlantECO) project, funded by European Union's Horizon 2020 Research and Innovation Program under Grant Agreement n. 862923.

**Author contributions:** Conceptualization: BBN, DI; Methodology: BBN; Investigation: BBN, DI; Visualization: BBN; Funding acquisition: BBN, DI; Project administration: BBN, DI; Supervision: BBN, DI; Writing – original draft: BBN; Writing – review & editing: BBN, DI

**Competing interests:** The authors declare that they have no competing interests.

**Data and materials availability:** The data used in this work are freely available through the European Union Copernicus Marine Service and Copernicus Climate Change Service data stores. All DOIs/links to the specific data products are provided in the supplementary materials. The Autoregressive Model fitting code is made available by prof. Tapio Schneider on GitHub (https://github.com/tapios/arfit)

# Figures and Tables

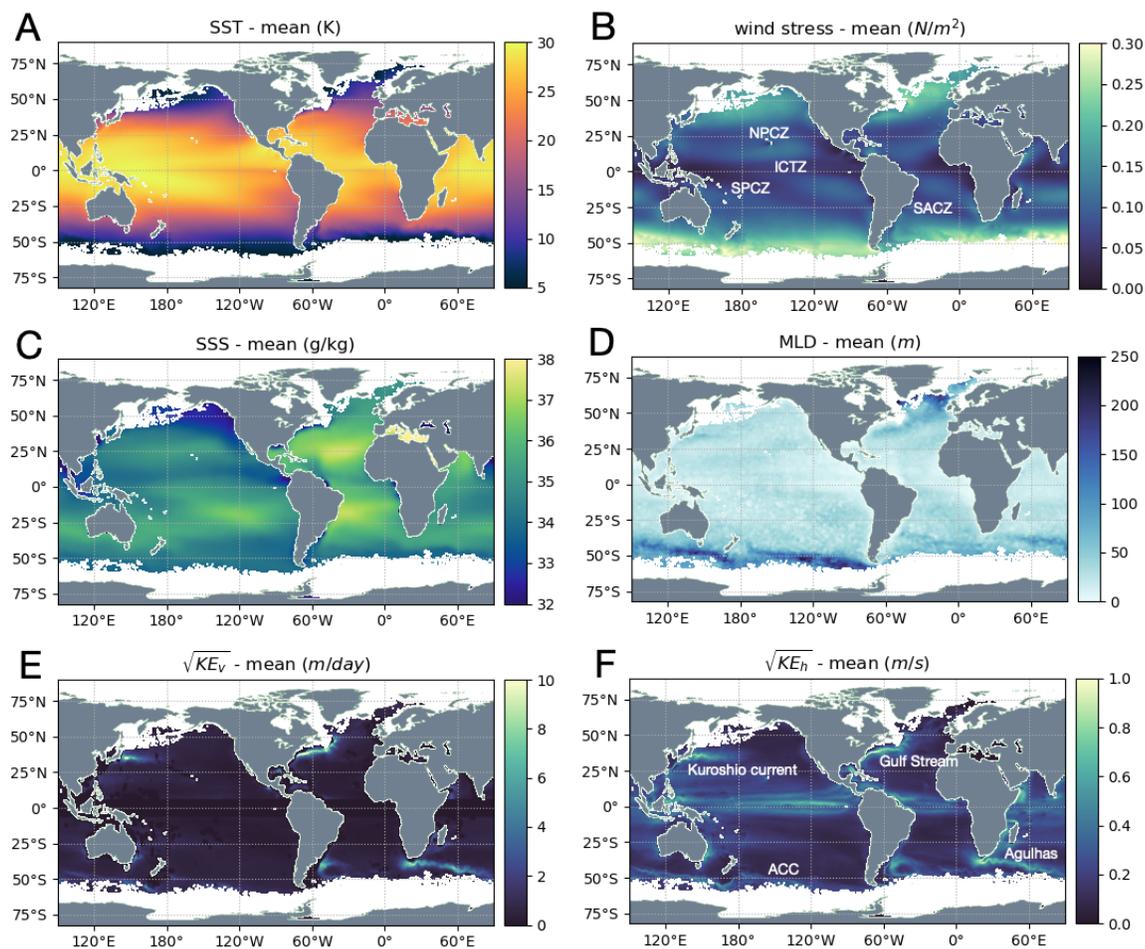

**Fig. 1. Observation-based dynamical ocean state variables averaged over the 1993-2018 period.** Sea surface temperature (**A**), wind stress intensity (**B**) sea surface salinity (**C**), upper mixed layer depth (**D**), intensity of the vertical exchanges at 100 m depth (**E**), intensity of the horizontal currents in the upper layer (**F**). (**E**) and (**F**) are both defined as the square root of corresponding kinetic energy components. In panel (**B**), four atmospheric convergence zones are indicated by their respective acronyms: Inter-Tropical Convergence Zone (ITCZ), North and South Pacific Convergence Zones (NPCZ, SPCZ), South Atlantic Convergence Zone (SACZ). In panel (**F**), the four main current systems are indicated.

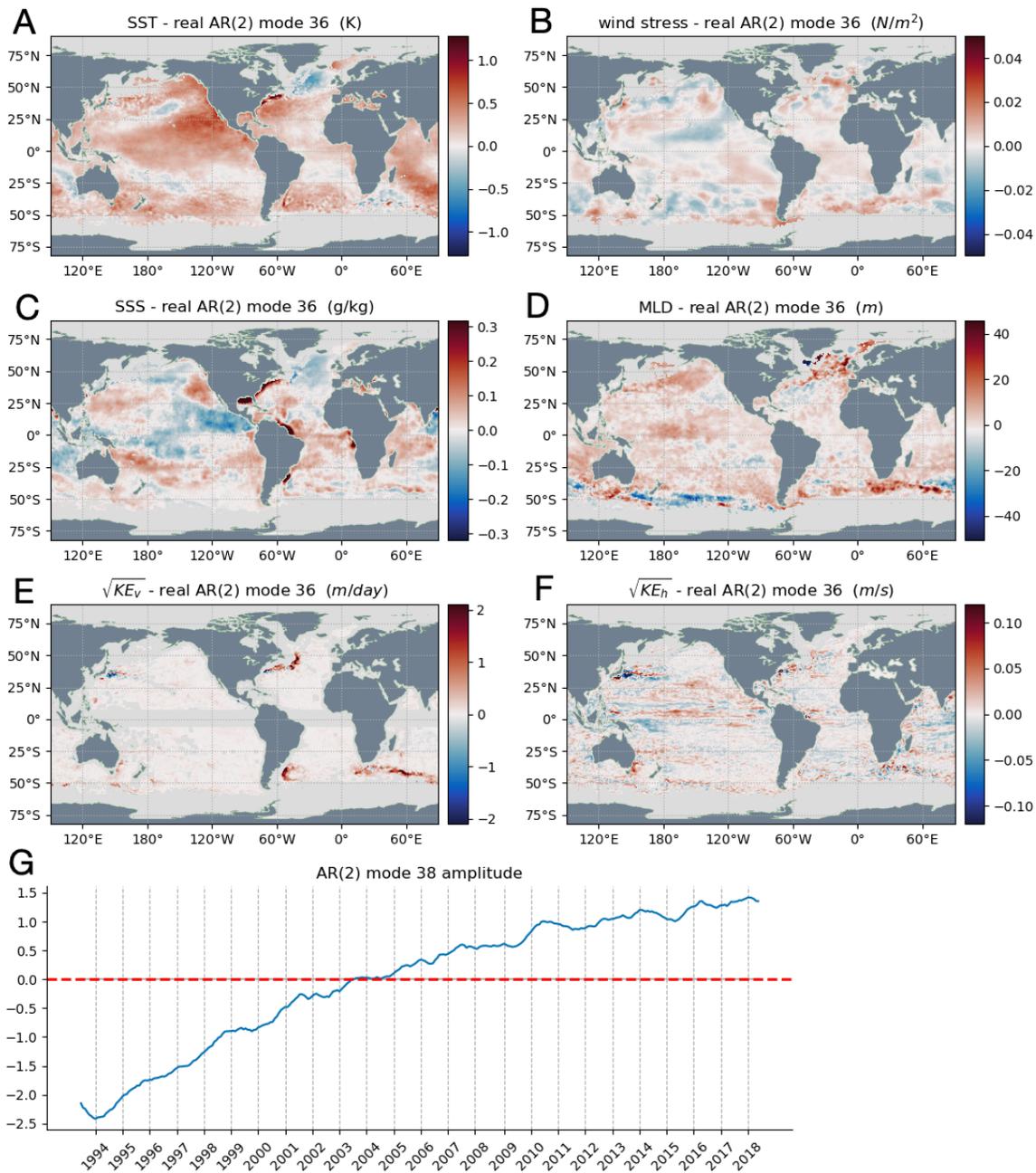

**Fig. 2**. **Patterns and amplitude of the POP trend mode estimated from observation-based dynamical ocean state variables.** Sea surface temperature (**A**), wind stress intensity (**B**) sea surface salinity (**C**), upper mixed layer depth (**D**), intensity of the vertical exchanges at 100 m depth (**E**), intensity of the horizontal currents in the upper layer (**F**). (**E**) and (**F**) are both defined as the square root of corresponding kinetic energy components. (**G**)Temporal amplitude of the trend mode.

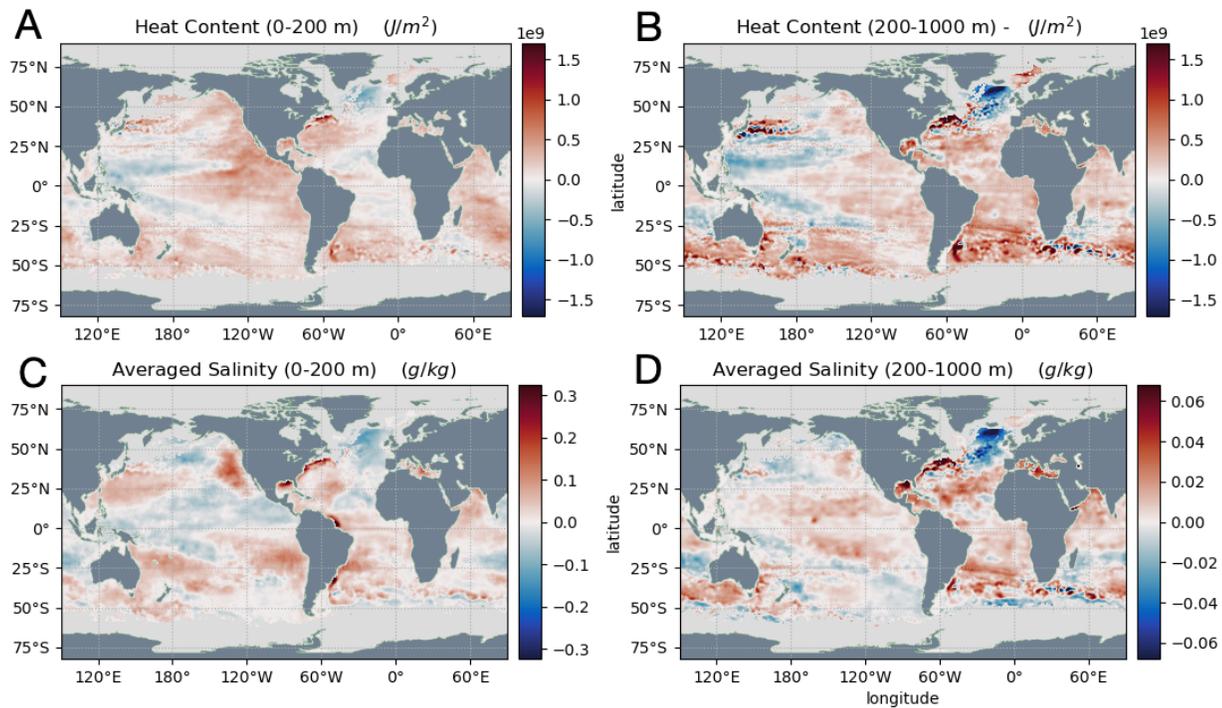

**Fig 3. Patterns of heat content and averaged salinity projected trends.** The trends have been projected on the heat data integrated over the two layers 0-200 m (**A**) and 200-1000 m (**B**), and on the salinity data averaged over the same layers (**C**, **D**).

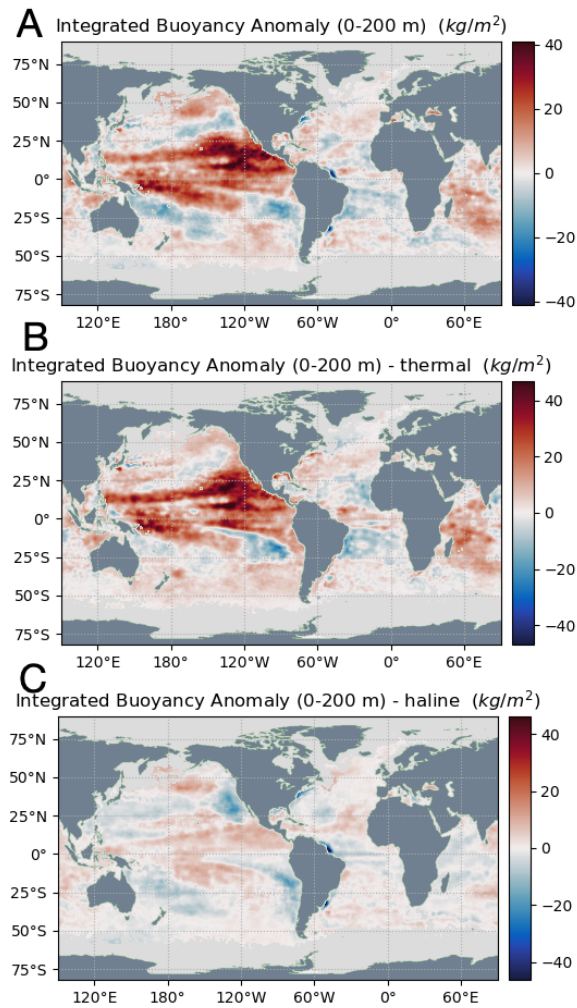

**Fig. 4. Patterns of integrated buoyancy anomaly projected trends**. The integrated buoyancy anomaly content is defined as the integral over depth of the density difference between each depth and the reference level (here 200 m). Total anomaly (**A**), thermal contribution to the buoyancy anomaly content (**B**), haline contribution to the buoyancy anomaly content (**C**).

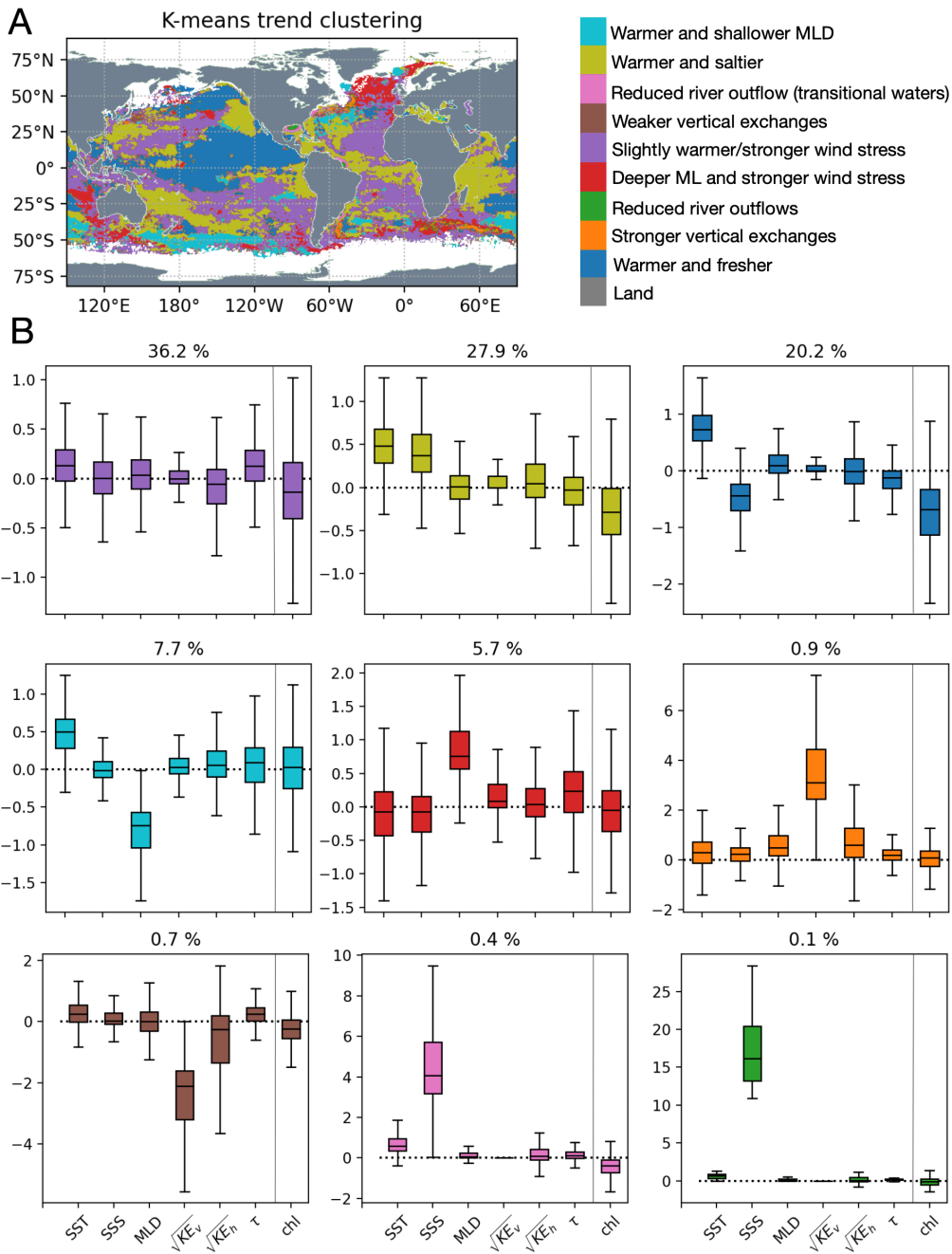

**Fig. 5**. **K-means clustering of trend patterns estimated from observation-based dynamical ocean state variables.** Spatial distribution of the clusters (**A**). Box plot of characteristic distribution of the state variables for each of the identified clusters (**B**). Boxes span from the lower quartile to the upper quartile, and lines represent the median values. Whiskers emanate from each box, illustrating the data's overall range. Values for chlorophyll-a are obtained from projected trends and are not used to identify the clusters.

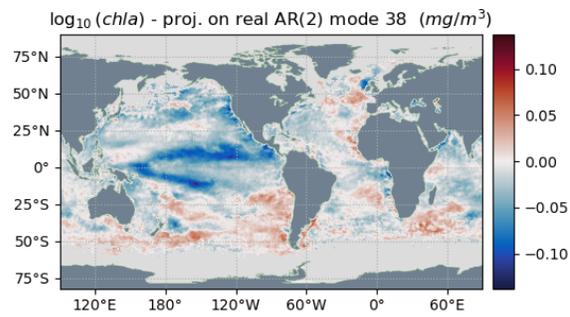

**Fig. 6. Patterns of chlorophyll-a projected trends**. Chlorophyll-a variable is preliminarily transformed through a logarithm to better resolve its dynamical range.

**Supplementary Materials**

  Supplementary Text
  Figs. S1 to S15
  Table S1

**Supplementary Text**

Principal Oscillation Patterns characterization

Ordering the POP oscillating modes by their characteristic period and analysing their spatial patterns, we can easily recognize previously identified physical processes and observe their signature in the different dynamical components (Table S1, Fig.S6-S11). In fact, some of these modes are characterized by relatively low frequencies. As such, given the reduced length of our time series, we have taken as a 'longer-term' reference the univariate analyses of the SST carried out and discussed by Di Lorenzo and colleagues(*19*). Despite the relatively short time series available, by comparing spatial patterns and related amplitudes, we could indeed show how our oscillating AR modes reflect mechanisms already identified starting from longer time series. Interestingly, even if most of these processes are dominated by anomalous temperature (and salinity patterns) in the Pacific tropical band, multivariate POP analysis clearly associates these signals also to extratropical variations in the other variables, both inside and outside the Pacific Ocean.

For the sake of simplicity, we have named each eigenmode focusing on its resemblance with the SST patterns and amplitudes previously described in the literature: the POP trend mode, three Pacific Decadal Variability (PDV) modes, i.e. a North Pacific-central Pacific mode (NP-CP, Fig.S7), a PDV decay mode (Fig.S8), a Kuroshio-Oyashio Extension mode (KOE, Fig.S9), and two El Niño-Southern Oscillation (ENSO) modes: an Eastern Pacific mode (EP, Fig.S10) and a central Pacific one (CP, Fig.S11). Additional modes whose patterns clearly resemble ENSO showed much lower excitations and are not shown here.

We estimated the linear regression slopes from each POP amplitude and multiplied it by the mean of the corresponding pattern to get the contribution of each mode to the total linear trend (see also Table S1). Summing the contributions from all 40 POP modes, we get a value of 0.0148 K/yr, which is consistent with previous similar estimates based on the same data(*26*), but strongly affected by the decreasing amplitude of the PDV decay mode over the period considered. The non-oscillating trend mode actually attains around 0.0221 K/yr.

North Atlantic trends reveal changes in the subpolar gyre circulation

The North Atlantic "warming hole", namely a marked SST cooling signal observed in the subpolar latitudes over the past century, has been often explained as the consequence of a slowdown of the Atlantic Meridional Overturning Circulation (AMOC)(*67*). The AMOC is known to play a special role in the climate system, due to its important function in exchanging heat and carbon between the upper layers and the deep ocean along its northward pathway, and its behaviour and projected impacts are quite debated. Recent modelling exercises indicate that approximately 50% of the observed cooling trend in the subpolar North Atlantic can be attributed to atmospheric factors alone(*68,69*). We observe that the cooling trend is linked to a strengthening of the westerly winds, which probably results in more heat being transferred from the ocean through turbulent heat exchanges. This change also significantly alters both the surface salinity distribution and the key currents within the subpolar gyre and North Atlantic Current system, as illustrated in Fig.S12. The subpolar North Atlantic's salinity stratification is crucial for sustaining the AMOC, and any significant surface freshening could anticipate a potential slowdown in the overturning(*70*). The wind-driven transport of freshwater from the Arctic acts as a link between the conditions in the Northwest Atlantic shelf and slope region and those of the eastern subpolar basins. In fact, the Labrador Current receives water from both the West Greenland Current and the Arctic, which are then directed towards the Labrador Shelf. This current has two distinct branches: one near the shore that flows along the Labrador Shelf, and another farther offshore along the Labrador shelf-break. As it reaches the tip of the Grand Banks, the current bifurcates into two directions: one retroflexes to the northeast and merges with the North Atlantic Current (NAC), while the other portion continues westward along the shelf. The retroflection occurs as part of a larger system of circulation adjustments within the subpolar gyre. These adjustments are influenced by wind patterns and have consequences for the strength of the Labrador Current and the position of the Gulf Stream(*71*). Both the subpolar and subtropical gyres display clear circulation changes (Fig. S12K): the horizontal current intensity trend highlights a northward displacement of the Gulf Stream and a successive widening of the surface transport around 60°W. Successively, the Gulf Stream divides into different branches that are undergoing marked changes: its southernmost branch feeds a moderately strengthening Azores current (that also appears to have shifted to the South), while its northern branch is shifted to the North of 50°N. This could be partly due to a shift/intensification of the wind patterns in the western North Atlantic (Fig. S12G), that accelerates the Labrador Current(*71*). Indeed, multiple studies have suggested that interactions with eddies and meanders of the Gulf Stream/North Atlantic Current (NAC) could cause the Labrador Current to be diverted away from the coastline. These interactions have been increasing over time, also leading to stronger vertical exchanges offshore the Grand Banks (Fig. S12I). Moreover, a reduction of the westward freshwater transport along the Scotian shelf, downstream the tip of the Grand Banks, contributes to the positive salinity trend, and to the marked salinity freshening in the area where the Gulf Stream merges with the Labrador current (Fig. S12C). Remarkably, all these changes drive a dipolar response in the MLD trend pattern, with a significant shoaling observed in the Labrador Sea and a concurrent deepening in the Irminger Sea. These changes are expected to have a significant impact on the formation rates and characteristics of subpolar mode waters, thus affecting the entire thermohaline circulation(*72*).

Comparing K-means and CLARA clusters

We show here the impact of the choice of the clustering algorithm, by comparing the results obtained with *K-means* (fig. 5) with those obtained with *CLARA* (Fig.S13). We also show the impact of increasing the number of clusters from 9 and to 13 (Fig.S14-S15).

When considering 9 clusters, *K-means* and *CLARA* identify basically the same dynamical regimes (and related Chl-a response) for red, blue, cyan, orange, green, olive and violet clusters. The differences are related, on one hand, to the percentage of the ocean surface that is associated with each regime, but also to the fact that a few regimes appear to be split into sub-regimes depending on the algorithm.

In particular, *K-means* also detects the small areas where a decrease in vertical exchanges is found (brown cluster) but splits the positive salinity trend extremes (green cluster in *CLARA*) in two different regimes (pink and green in *K-means*). Instead, *CLARA* describes a more gradual change between violet cluster and olive cluster. *CLARA* more clearly identifies increasing wind stress (violet cluster), and the stronger salinity signal of olive cluster, but introduces a cluster describing a slowdown of surface circulation (shown in light violet), and a sort of transition cluster depicted in light olive.

Increasing the number of clusters is not adding any fundamental information, but mostly splitting the identified regimes depending on the intensity of the associated responses. Specifically, we present in Fig.S14-S15 the maps and box plots obtained when considering 13 clusters. To facilitate the intercomparison, in all figures we kept the same colours for all clusters that basically identify the same core trend behaviours, and used for the new clusters light-colour variations whenever those could be clearly associated with similar responses (i.e. slight variations in the intensity of the related trends). Only few colours identify clusters with unique behaviours, but these are always linked to a quite small percentage of the surface, and the maps always display the same overall patterns. The different granularity results in more noisy patterns when a higher number of clusters is selected, and we believe that 9 clusters are a better choice to discuss the geography of trends.

In any case, we must stress that we are always calculating and discussing the distributions of characteristic trends for each variable within each cluster. The values provided in the box plots are scientifically sound and can be unambiguously discussed whatever the number of clusters considered.

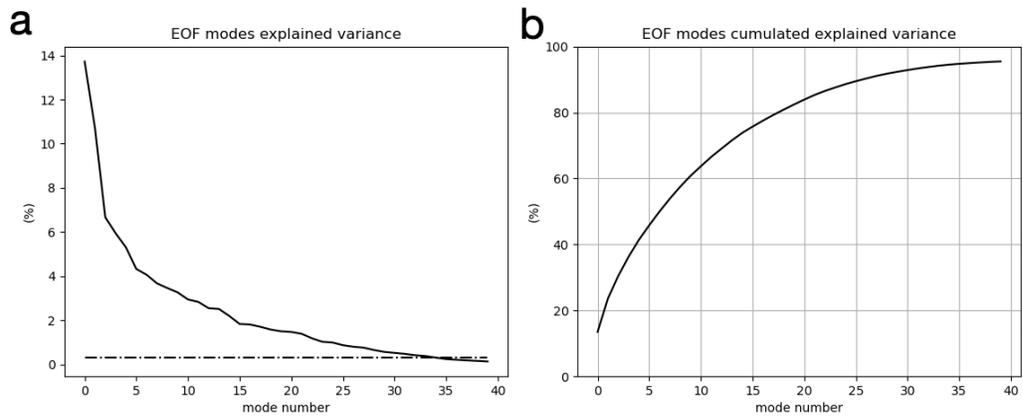

**Fig.S1.** (**a**) Percentage of variance and (**b**) percentage of cumulated variance of the EOF decomposition. In (**a**) the dot-dash line identifies the average percentage computed from all 326 EOF eigenvalues.

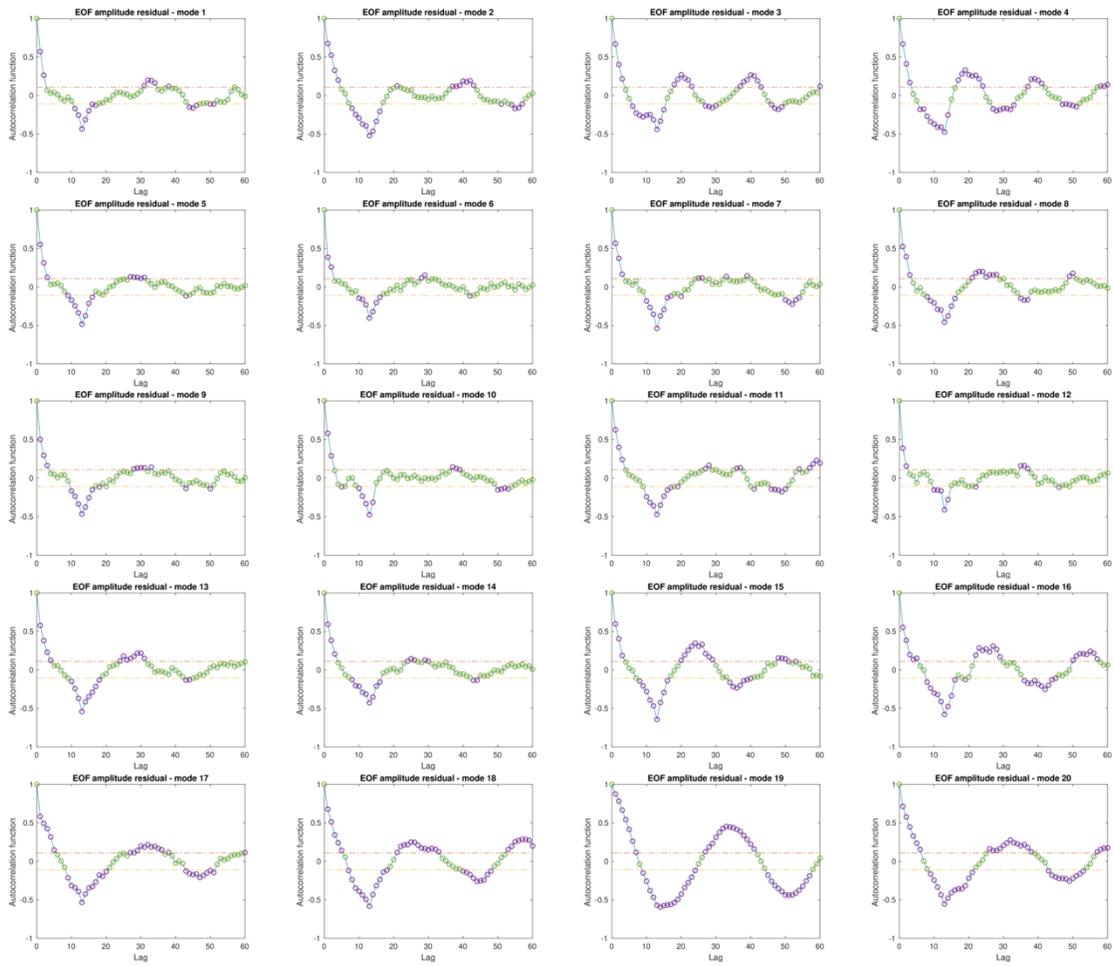

**Fig.S2.** Autocorrelation function of the AR(1) model residuals for each of the 20 PCs in input.

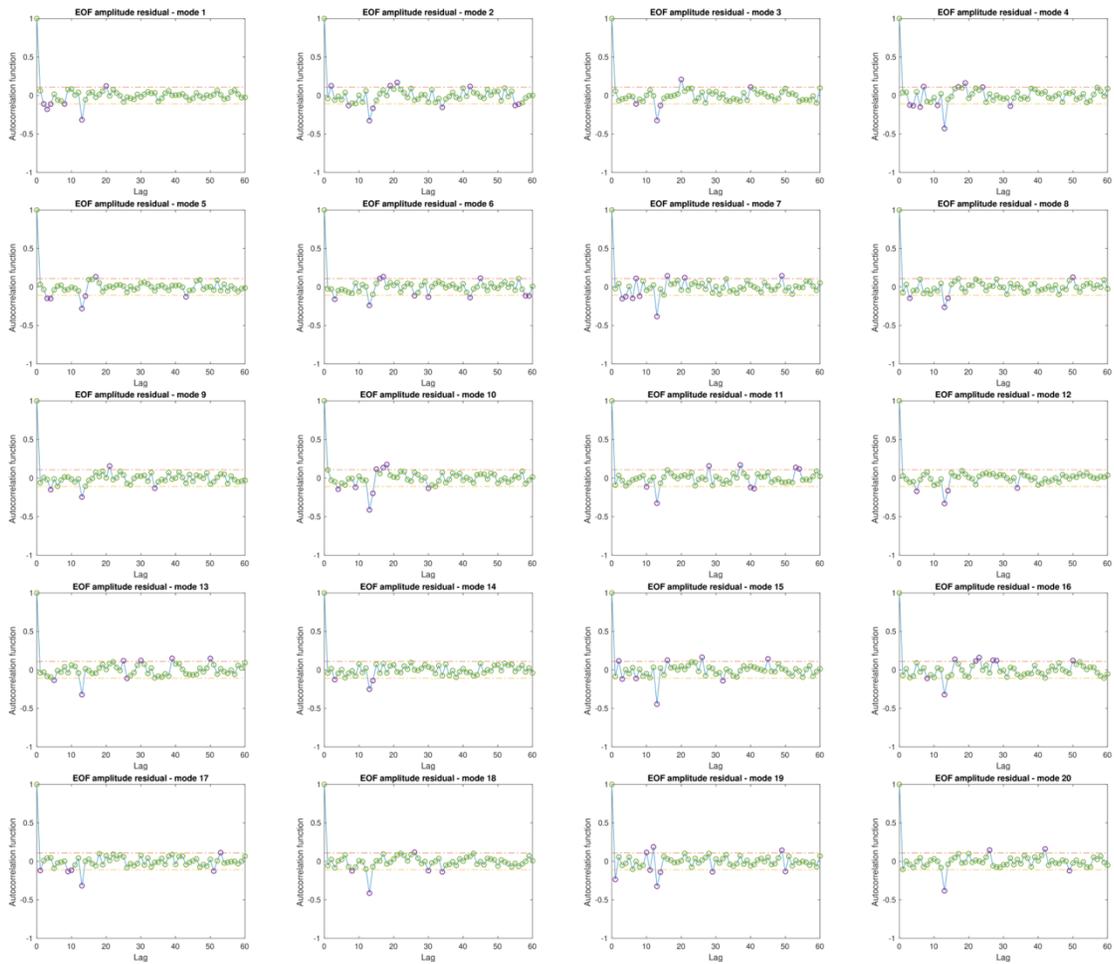

**Fig.S3.** Autocorrelation function of the AR(2) model residuals for each of the 20 PCs in input.

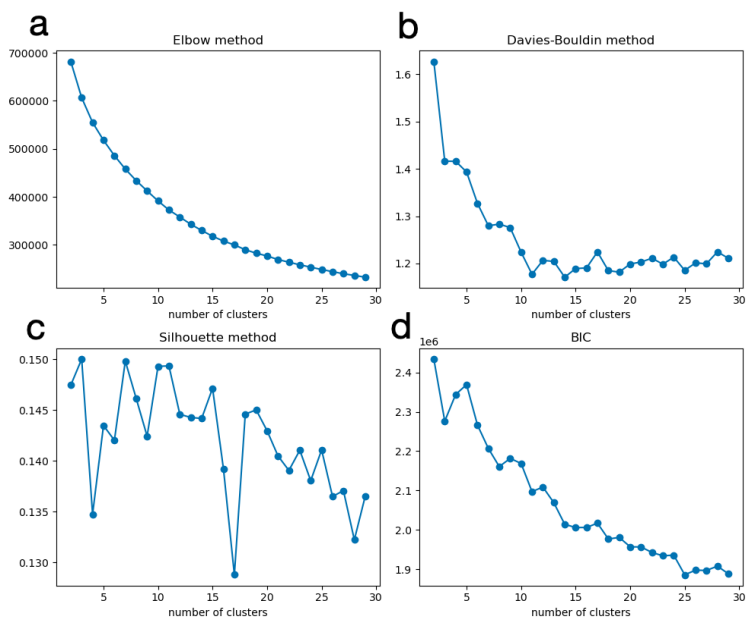

**Fig.S4. K-means clustering scores as a function of the number of clusters**: (**a**) within-cluster sum of squares ("elbow" criterion); (**b**) Davies-Bouldin Index; (**c**) Silhouette score; (**d**) Bayesian Information Criterion

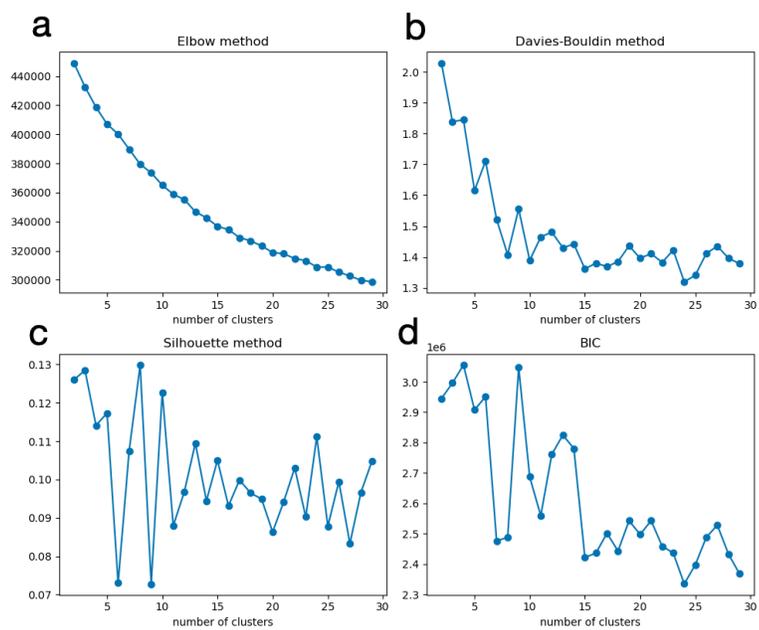

**Fig.S5.** *CLARA* **clustering scores as a function of the number of clusters**: (**a**) within-cluster sum of squares ("elbow" criterion); (**b**) Davies-Bouldin Index; (**c**) Silhouette score; (**d**) Bayesian Information Criterion

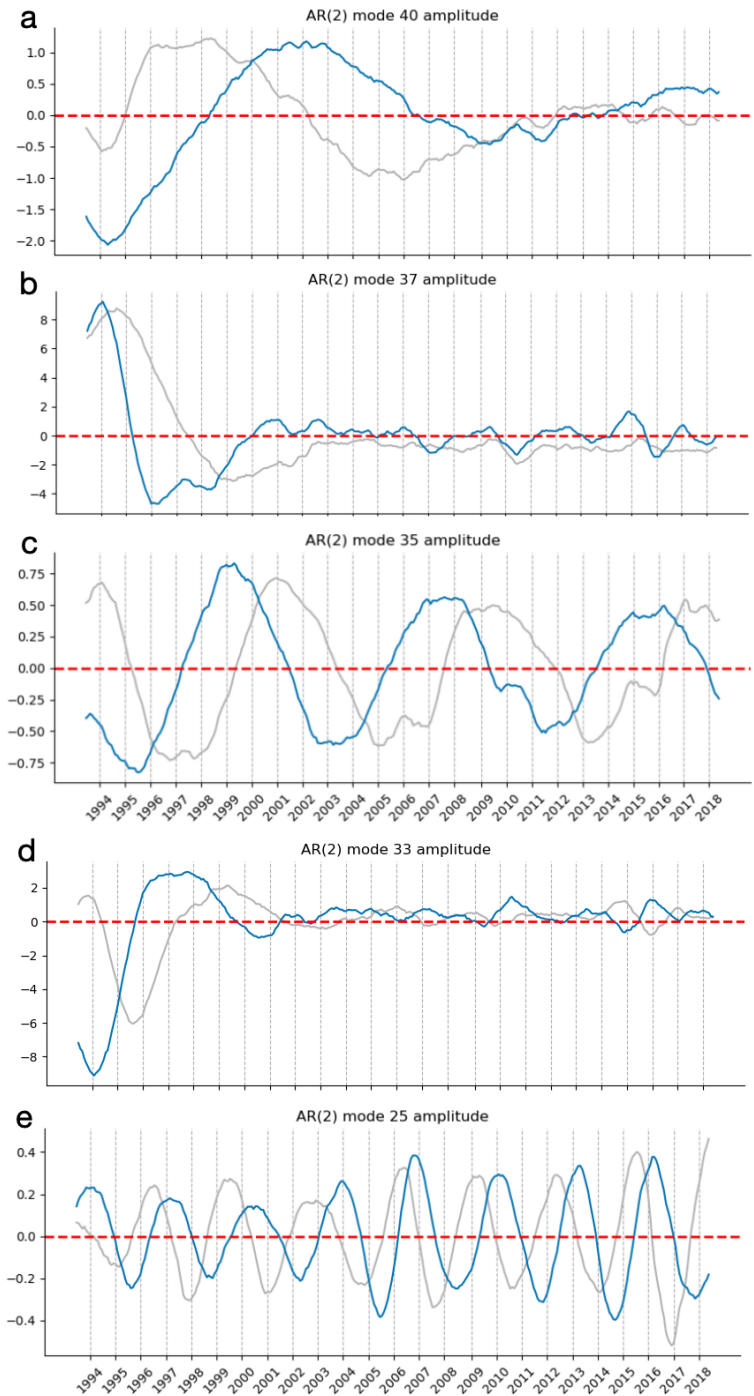

**Fig. S6. Real (blue line) and imaginary (light grey line) components of selected POP modes**: North Pacific-central Pacific mode (NP-CP) Pacific Decadal Variability (PDV) mode (**a**), PDV decay mode (**b**), PDV Kuroshio-Oyashio Extension (KOE) mode (**c**), Eastern Pacific (EP) El Niño-Southern Oscillation (ENSO) mode (**d**), central Pacific (CP) ENSO mode (**e**).

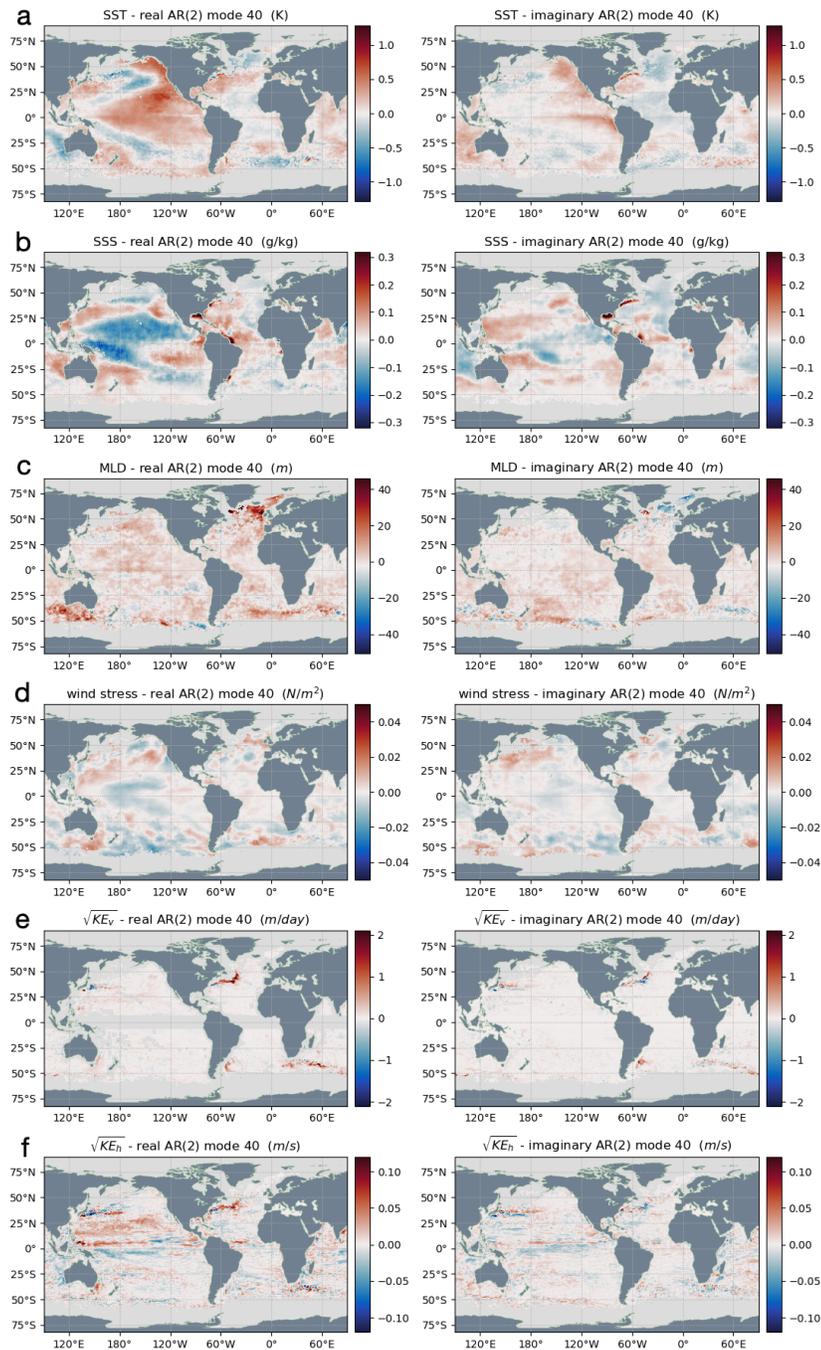

**Fig. S7. Real and imaginary components of the patterns of the North Pacific-central Pacific Decadal Variability (PDV NP-CP) POP mode**. Sea surface temperature (**a**), wind stress intensity (**b**) sea surface salinity (**c**), upper mixed layer depth (**d**), intensity of the vertical exchanges at 100 m depth (**e**), intensity of the horizontal currents in the upper layer (**f**).

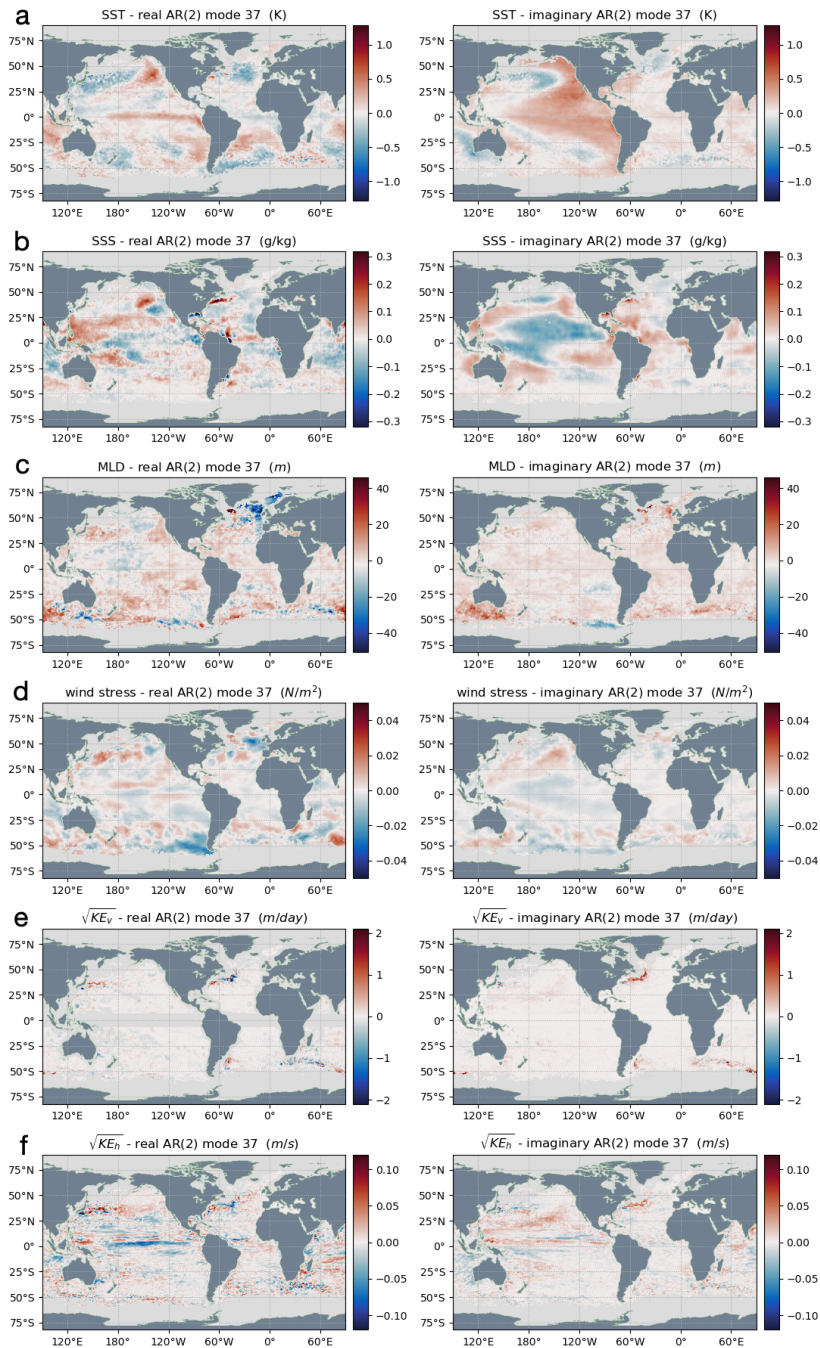

**Fig.S8. Real and imaginary components of the patterns of the Pacific Decadal Variability decay POP mode.** Sea surface temperature (**a**), wind stress intensity, (**b**) sea surface salinity (**c**), upper mixed layer depth (**d**), intensity of the vertical exchanges at 100 m depth (**e**), intensity of the horizontal currents in the upper layer (**f**).

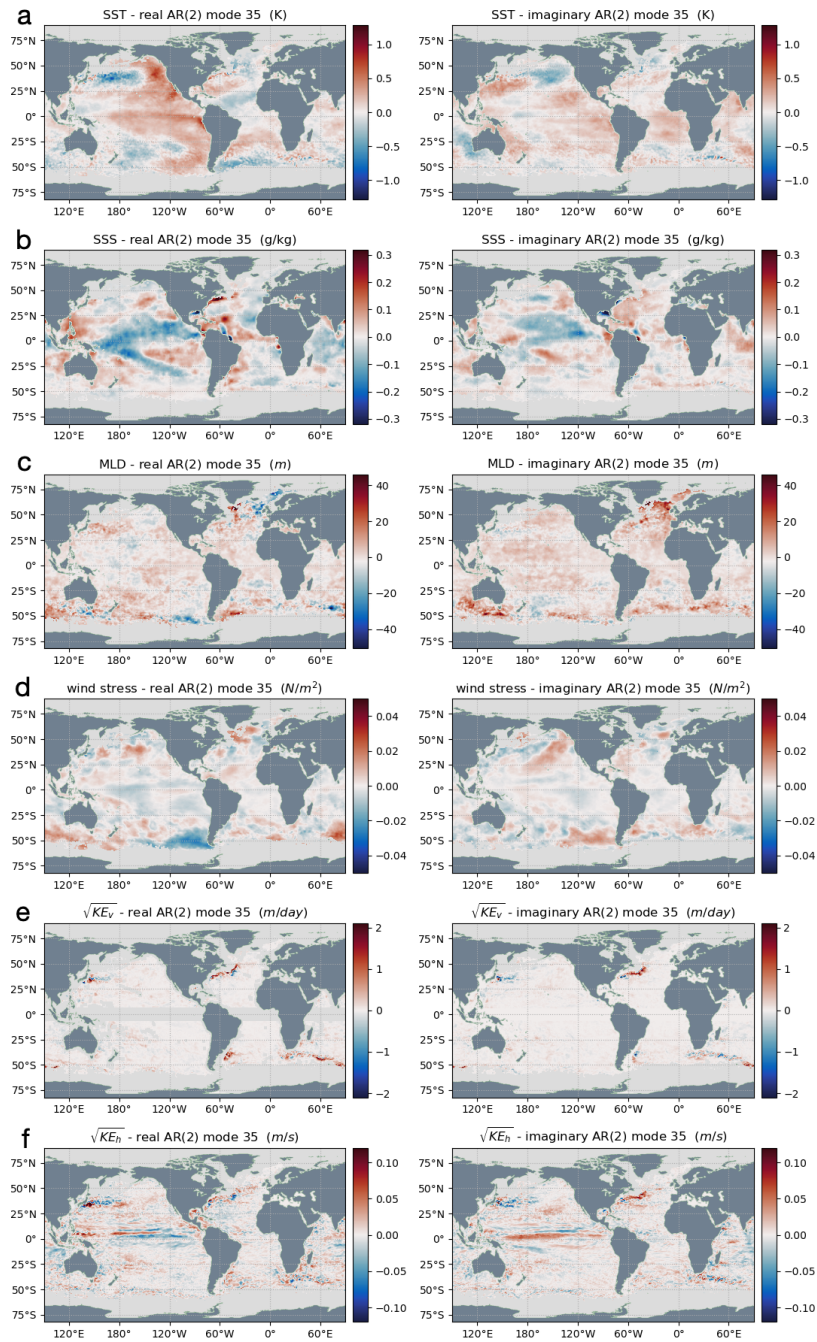

**Fig.S9. Real and imaginary components of the patterns of the Pacific Decadal Variability Kuroshio-Oyashio Extension (PDV-KOE) POP mode**. Sea surface temperature (**a**), wind stress intensity (**b**) sea surface salinity (**c**), upper mixed layer depth (**d**), intensity of the vertical exchanges at 100 m depth (**e**), intensity of the horizontal currents in the upper layer (**f**).

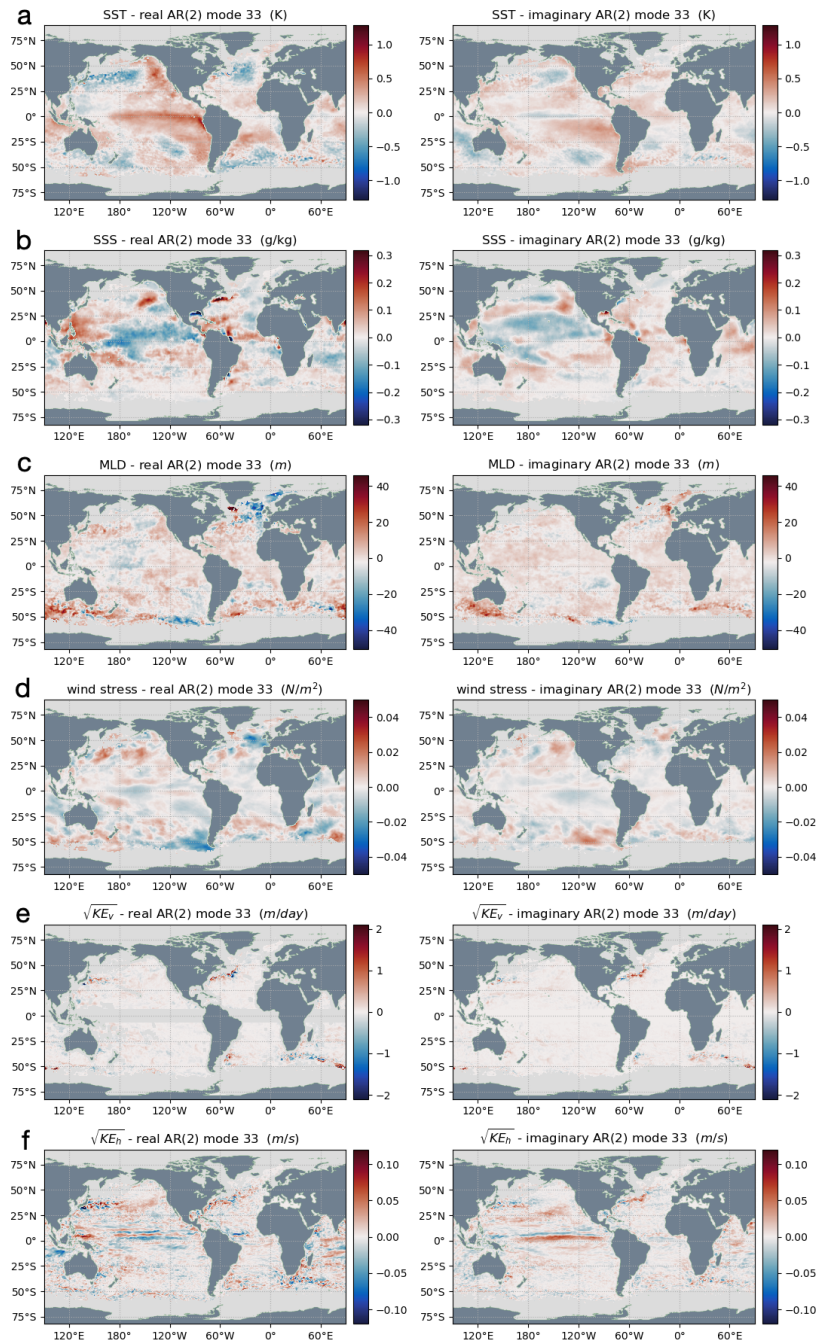

**Fig.S10. Real and imaginary components of the patterns of the Eastern Pacific (EP) El Niño-Southern Oscillation (ENSO) POP mode**. Sea surface temperature (**a**), wind stress intensity (**b**) sea surface salinity (**c**), upper mixed layer depth (**d**), intensity of the vertical exchanges at 100 m depth (**e**), intensity of the horizontal currents in the upper layer (**f**).

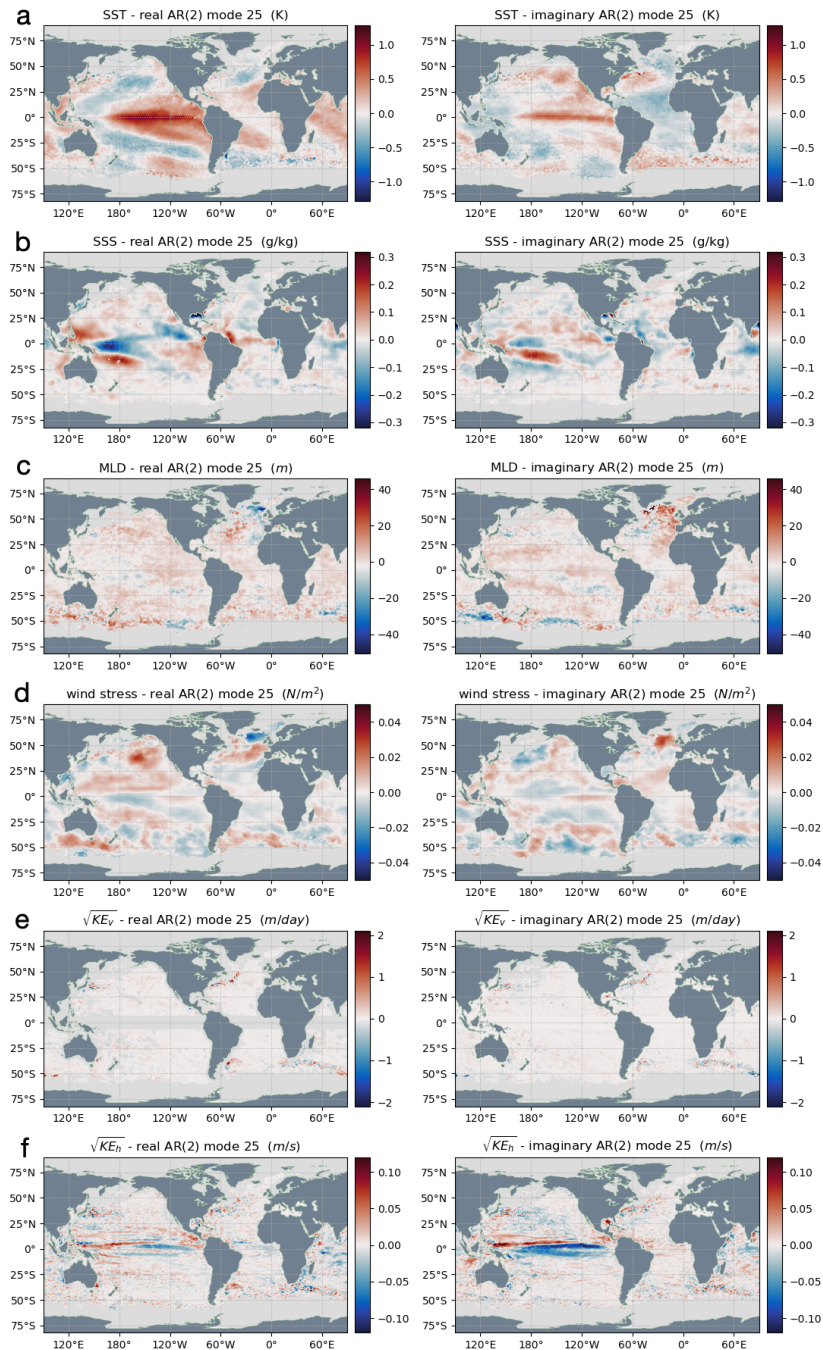

**Fig.S11. Real and imaginary components of the patterns of the Central Pacific (CP) El Niño-Southern Oscillation (ENSO) POP mode**. Sea surface temperature (**a**), wind stress intensity (**b**) sea surface salinity (**c**), upper mixed layer depth (**d**), intensity of the vertical exchanges at 100 m depth (**e**), intensity of the horizontal currents in the upper layer (**f**).

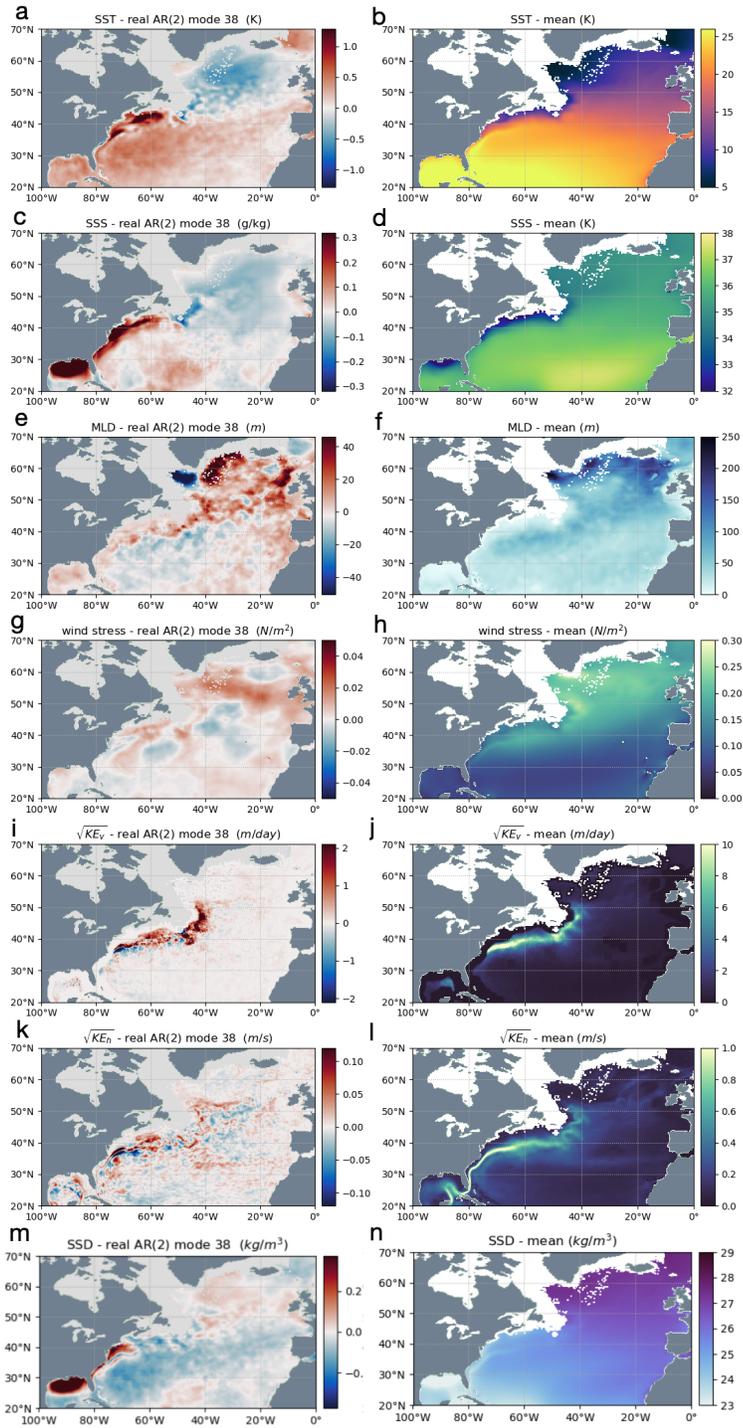

**Fig.S12**. **Zoom over the North Atlantic Ocean of the trend patterns and mean values estimated from the observation-based dynamical ocean state variables.** Sea surface temperature (**a**, **b**), wind stress intensity (**c**, **d**) sea surface salinity (**e**, **f**), upper mixed layer depth (**g**, **h**), intensity of the vertical exchanges at 100 m depth (**i**, **j**), intensity of the surface horizontal currents (**k**, **l**), sea surface density (**m**, **n**).

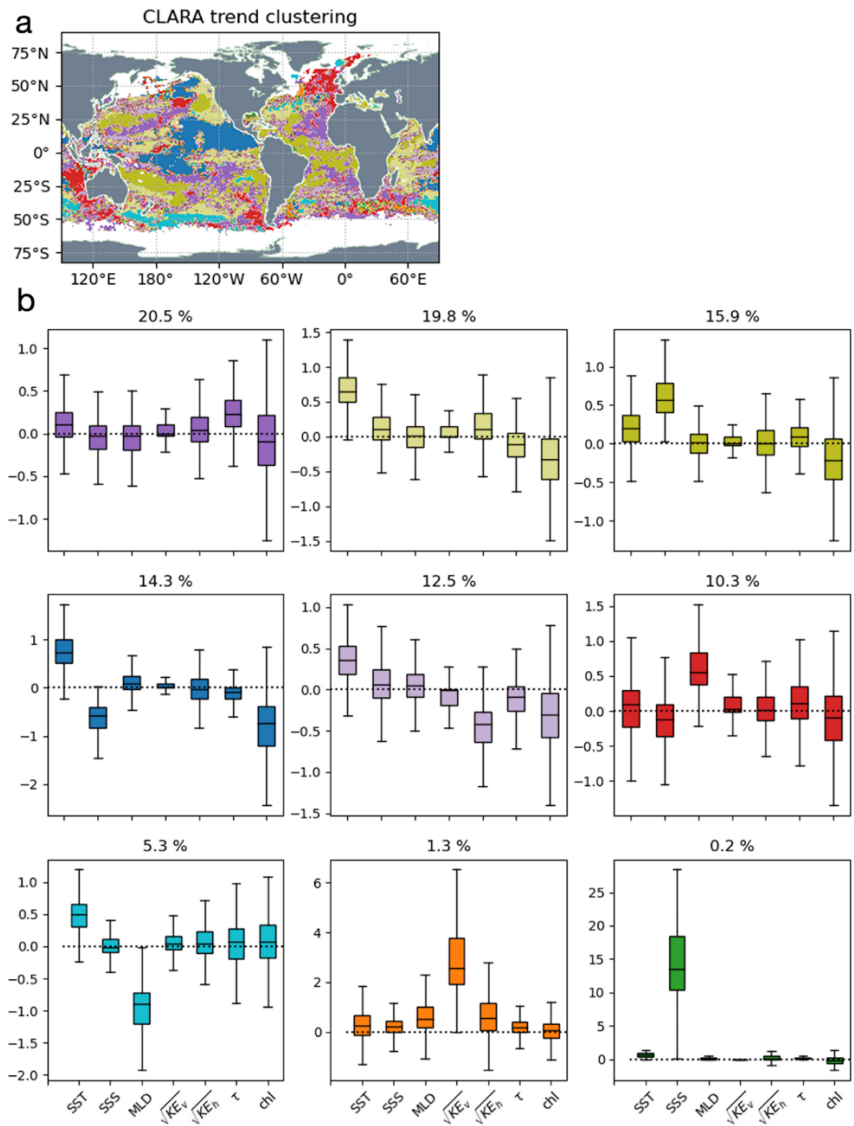

**Fig.S13. *CLARA* clustering of trend patterns estimated from observation-based dynamical ocean state variables, considering 9 clusters.** Spatial distribution of the clusters (**a**). Box plot of characteristic distribution of the state variables for each of the identified clusters (**b**). Boxes span from the lower quartile to the upper quartile, and lines represent the median values. Whiskers emanate from each box, illustrating the data's overall range. Values for chlorophyll-a are obtained from projected trends and are not used to identify the clusters.

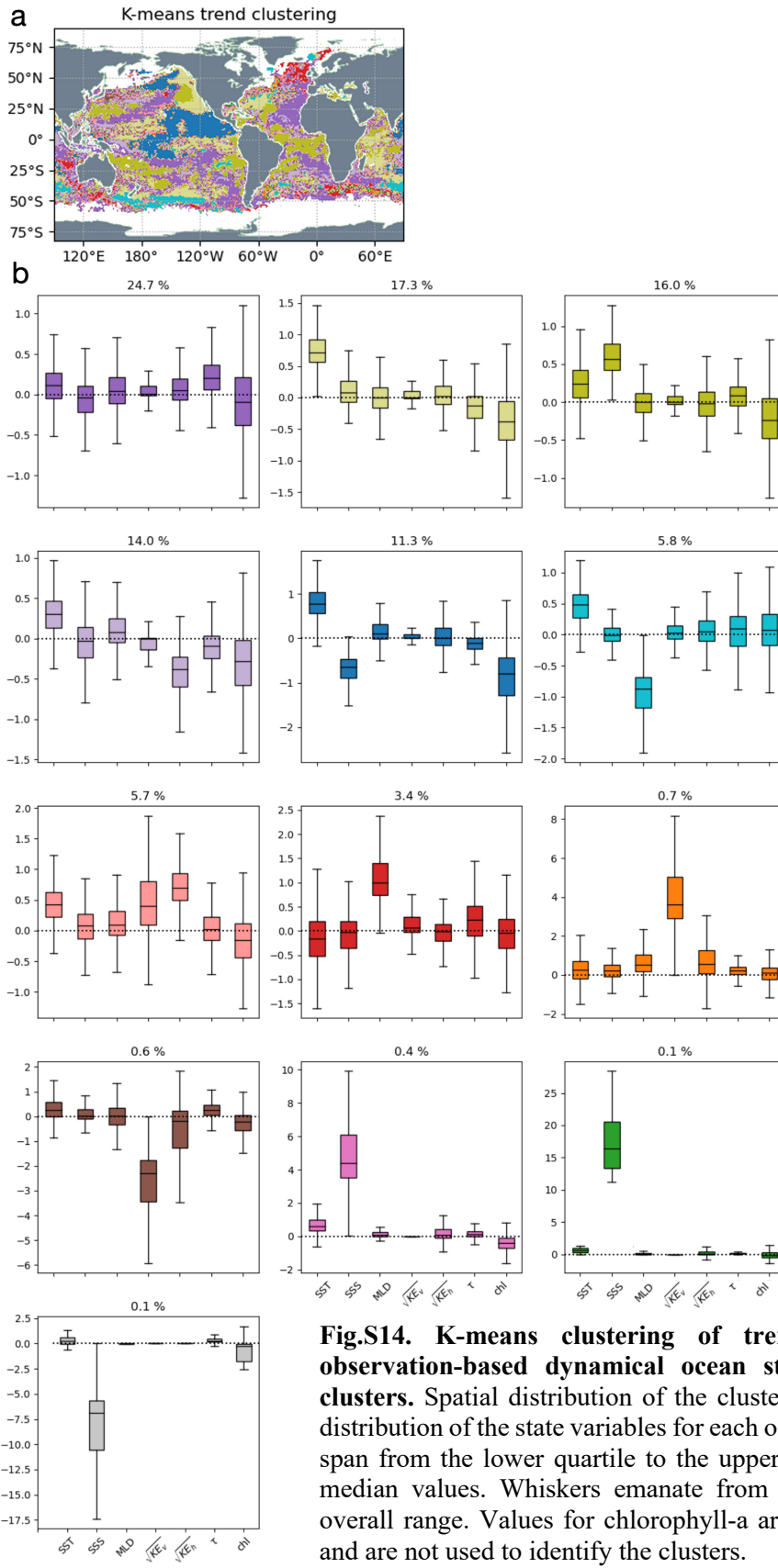

**Fig.S14. K-means clustering of trend patterns estimated from observation-based dynamical ocean state variables, considering 13 clusters.** Spatial distribution of the clusters (**a**). Box plot of characteristic distribution of the state variables for each of the identified clusters (**b**). Boxes span from the lower quartile to the upper quartile, and lines represent the median values. Whiskers emanate from each box, illustrating the data's overall range. Values for chlorophyll-a are obtained from projected trends and are not used to identify the clusters.

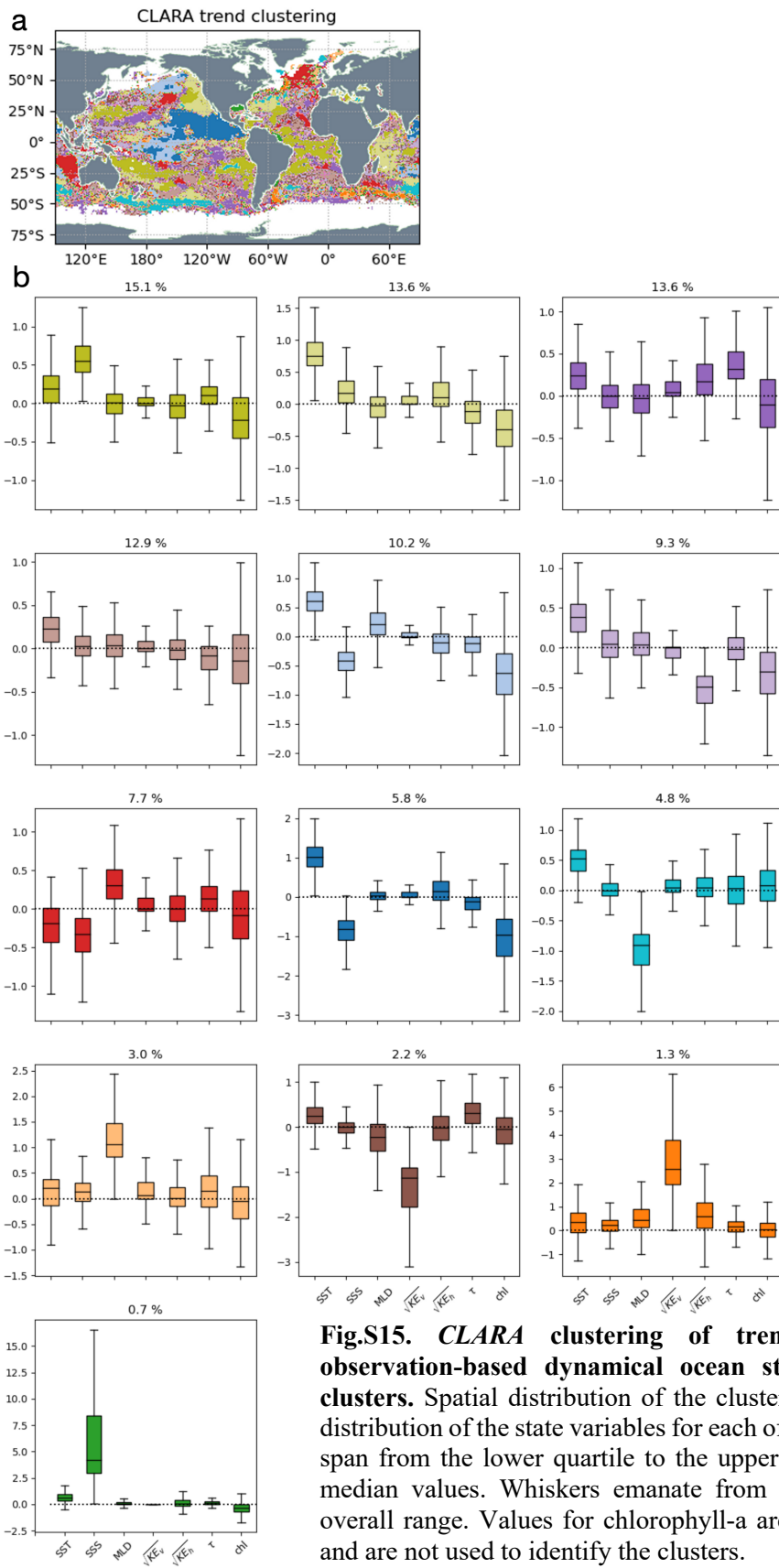

**Fig.S15. *CLARA* clustering of trend patterns estimated from observation-based dynamical ocean state variables, considering 13 clusters.** Spatial distribution of the clusters (**a**). Box plot of characteristic distribution of the state variables for each of the identified clusters (**b**). Boxes span from the lower quartile to the upper quartile, and lines represent the median values. Whiskers emanate from each box, illustrating the data's overall range. Values for chlorophyll-a are obtained from projected trends and are not used to identify the clusters.

| AR(2) mode | Period (years) | Damping scale (years) | Excitation | SST linear trend (K/yr) | Physical process |
|---|---|---|---|---|---|
| **38** | Inf | 12.8±21.8 | 1.0 | 0.0221±0.0006 (±0.002) | Trend |
| **(39) 40** | 17.3±9.0 | 10.1±16.0 | 2.77 | -0.0024±0.0004 | PDV (NP-CP) |
| **(36) 37** | 10.5±15.2 | 3.11±13.7 | 16.8 | -0.011±0.002 | PDV decay mode |
| **(34) 35** | 8.8±1.7 | 23.6±67.5 | 3.5 | -0.0006±0.0002 | PDV-KOE |
| **(32) 33** | 7.4±10.1 | 2.9±6.6 | 11.2 | 0.0039±0.0007 | EP-ENSO |
| **(24) 25** | 3.4±0.2 | 56.5±383 | 6.2 | 0.0002±0.0002 | CP-ENSO |

**Table S1.** Period, dumping scale (both including 95% confidence intervals), excitation, sea surface temperature linear regression slope (including 95% confidence intervals) of the empirically estimated eigenmodes and related physical process. For the trend mode, we also indicate in parenthesis the 95% confidence interval estimated from the AR(2) models based on 18 to 22 PCs (see Methods).